\documentstyle[12pt,epsfig]{article}
\begin{document}
\renewcommand{\theequation}{\arabic{section}.\arabic{equation}}
\setlength{\parskip}{0.45cm}
\setlength{\baselineskip}{0.75cm}
%
%
\begin{titlepage}
\begin{flushright}
DO-TH 95/13 \\ RAL-TR-95-042  \\ August 1995 \\
updated version (December 1995)
\end{flushright}
\vspace{1.0cm}
\begin{center}
\Large
\hbox to\textwidth{\hss
{\bf Next-to-Leading Order Radiative Parton Model Analysis} \hss}

\vspace{0.1cm}
\hbox to\textwidth{\hss
{\bf of Polarized Deep Inelastic Lepton Nucleon Scattering} \hss}

\vspace{1.5cm}
\large
M.\ Gl\"{u}ck, E.\ Reya, M.\ Stratmann\\
\vspace{0.5cm}
\normalsize
Universit\"{a}t Dortmund, Institut f\"{u}r Physik, \\
\vspace{0.1cm}
D-44221 Dortmund, Germany \\
\vspace{1.2cm}
\large
W. Vogelsang \\
\vspace{0.5cm}
\normalsize
Rutherford Appleton Laboratory \\
\vspace{0.1cm}
Chilton, Didcot, Oxon OX11 0QX, England \\
\vspace{1.5cm}
{\bf Abstract}
\vspace{-0.3cm}
\end{center}
A next-to-leading order QCD analysis of spin asymmetries and structure
functions in polarized deep inelastic lepton nucleon scattering is
presented within the framework of the radiative parton model. A consistent
NLO formulation of the $Q^2$-evolution of polarized parton distributions
yields two sets of plausible NLO spin dependent parton distributions in
the conventional $\overline{\rm{MS}}$ factorization scheme. They respect
the fundamental positivity constraints down to the low resolution
scale $Q^2=\mu^2_{NLO}=0.34\,{\rm{GeV}}^2$. The $Q^2$-dependence of the
spin asymmetries $A_1^{p,n,d}(x,Q^2)$ is similar to the leading-order (LO)
one in the range $1\le Q^2\le 20\,{\rm{GeV}}^2$ and is shown to be
non-negligible for $x$-values relevant for the analysis of the present
data and possibly forthcoming data at HERA.
\end{titlepage}
\section{Introduction}
Recently, a leading order (LO) QCD analysis of polarized deep inelastic
lepton nucleon scattering has been performed \cite{ref1}
within the framework of the radiative parton model.
The first moments $\Delta f(Q^2)$ of polarized parton distributions
$\delta f(x,Q^2)$,
\begin{equation}
\Delta f(Q^2) \equiv \int_0^1 dx\; \delta f(x,Q^2)\;\;\;,
\end{equation}
where $f=u,\bar{u},d,\bar{d},s,\bar{s},g$, were subject to two very
different sets of theoretical constraints related to two different views
concerning the flavor $SU(3)$ [$SU(3)_f$] symmetry properties of hyperon
$\beta$-decays. One set ('standard' scenario) assumed an unbroken
$SU(3)_f$ symmetry between the relevant matrix elements while the other
set ('valence' scenario) assumed an extremely broken $SU(3)_f$ symmetry
reflected in relating \cite{ref2} the hyperon $\beta$-decay matrix
elements to the first moments of the corresponding {\em{valence}}
distributions. Both scenarios gave satisfactory descriptions \cite{ref1}
of the measured spin-asymmetries [3-9]
$A_1^{p,n}(x,Q^2) \simeq g_1^{p,n}(x,Q^2)/F_1^{p,n}(x,Q^2)$, although
the polarized gluon density $\delta g(x,Q^2)$, which enters in LO
only via the $Q^2$-evolution equations, was only weakly constrained by
present data. The total helicity carried by quarks
\begin{equation}
\Delta \Sigma (Q^2)\equiv \sum_{q=u,d,s}
(\Delta q(Q^2)+\Delta \bar{q}(Q^2))\;\;\;,
\end{equation}
which is $Q^2$-independent in LO, turned out to be $\Delta \Sigma \simeq
0.3$ in both scenarios with an average total gluonic helicity
$\Delta g(Q^2=4\,{\rm{GeV}}^2) \simeq 1.5$. A specific feature of our radiative
LO analysis is that the polarized leading twist parton densities
$\delta f(x,Q^2)$ are valid down to $Q^2=\mu_{LO}^2\simeq 0.23\,{\rm{GeV}}^2$
and that the fundamental positivity constraints
\begin{equation}
\left| \delta f(x,Q^2)\right| \le f(x,Q^2)
\end{equation}
are respected down to this low resolution scale $Q^2=\mu_{LO}^2$ as well.
The 'standard' scenario requires a finite total
strange sea helicity of $\Delta s =\Delta \bar{s} \simeq -0.05$ in order
to account for a reduction of $\Gamma_1^p$ with respect to the
Gourdin and Ellis and Jaffe estimate \cite{ref10}
\begin{equation}
\Gamma_{1,EJ}^p=\frac{1}{12} (F+D) + \frac{5}{36} (3F-D) \simeq 0.185
\end{equation}
where
\begin{equation}
\Gamma_1^p(Q^2) \equiv \int_0^1 dx\; g_1^p(x,Q^2)\;\;\;.
\end{equation}
Within the 'valence' scenario, on the contrary, a  negative light sea
helicity $\Delta \bar{u}=\Delta \bar{d} \equiv \Delta \bar{q}
\simeq -0.07$ suffices ($\Delta s = \Delta \bar{s} =0$) for reducing
$\Gamma_{1,EJ}^p$. Furthermore, in both scenarios we \cite{ref1}
predict $\Gamma_1^p\simeq 0.15$ and $\Gamma_1^n\simeq -0.06$ in
agreement with recent experiments [5-9], with the Bj\o rken sum rule
being manifestly satisfied.

While our LO analysis was being completed, a full next-to-leading order
(NLO) calculation of all polarized two-loop splitting functions
$\delta P_{ij}^{(1)}(x),\;i,j=q,g$, in the conventional
$\overline{\rm{MS}}$ factorization scheme has appeared for the first
time \cite{ref11}. It is the purpose of this article to present first
a consistent NLO formulation of spin-dependent parton distributions,
making use of the NLO results of ref.\cite{ref11}, in particular for
(Mellin) $n$-moments of structure functions and parton densities where
the solutions of the NLO evolution equations can be obtained
analytically. Using these formal results we then proceed to perform
a quantitative NLO analysis of $A_1^{p,n}(x,Q^2)$ and $g_1^{p,n}(x,Q^2)$,
and will present two sets of NLO $\delta f(x,Q^2)$ for the two
scenarios discussed at the beginning. Since most NLO analyses concerning
unpolarized hard processes and parton distributions have been performed
in the $\overline{\rm{MS}}$ factorization scheme, it is convenient to remain
within this factorization scheme also for polarized hard processes and
spin-dependent parton distributions. This is particularly relevant
for the parton distributions which have to satisfy the
fundamental positivity constraints (1.3) at any
value of $x$ and scale $Q^2$, as calculated by the unpolarized and
polarized evolution equations, within the {\em{same}}
factorization scheme.
In addition we also repeat our previous LO analysis \cite{ref1} since new
data have been published very recently \cite{ref6,ref7}.
\section{NLO Parton Distributions and their
\protect{$\protect{\bf{Q^2}}$}-Evolution}
\setcounter{equation}{0}
Measurements of polarized deep inelastic lepton nucleon scattering yield
direct information [3-9,12] on the spin-asymmetry
\begin{equation}
A_1^N (x,Q^2) \simeq \frac{g_1^N (x,Q^2)}{F_1^N (x,Q^2)}
=\frac{g_1^N(x,Q^2)}{F_2^N (x,Q^2)/ \left[ 2x(1+R^N(x,Q^2)) \right] } \:\:\: ,
\end{equation}
$N=p,n$ and $d=(p+n)/2$, where in the latter case we have used
$g_1^d=\frac{1}{2} (g_1^p+g_1^n) (1-\frac{3}{2} \omega_D)$ with
$\omega_D=0.058$ \cite{ref7,ref9};
$R \equiv F_L/2xF_1 =(F_2-2 x F_1)/2xF_1$ and
subdominant contributions have, as usual, been neglected.
In NLO, $A_1^N(x,Q^2)$ is related to the polarized ($\delta f^N$) and
unpolarized ($f^N$) quark and gluon distributions in the following way:
\begin{eqnarray}
\nonumber
g_1^N(x,Q^2) &=& \frac{1}{2} \sum_q e_q^2\; \Bigg\{ \delta q^N(x,Q^2)+
\delta \bar{q}^N(x,Q^2)+\\
&+& \frac{\alpha_s(Q^2)}{2\pi} \left[ \delta C_q * \left( \delta q^N+
\delta \bar{q}^N\right) +\frac{1}{f} \delta C_g * \delta g\right]
\Bigg\}
\end{eqnarray}
with the convolutions being defined by
\begin{equation}
(C*q) (x,Q^2) =\int_x^1 \frac{dy}{y}\;C\!\left(\frac{x}{y}\right) q(y,Q^2)
\end{equation}
and where the appropriate spin-dependent Wilson coefficients in the
$\overline{\rm{MS}}$ scheme are given by (see \cite{ref11}, for example,
and references therein)
\begin{eqnarray}
\nonumber
\delta C_q(x) &=& C_F \Bigg[(1+x^2) \left(\frac{\ln (1-x)}{1-x}
\right)_{\!\!+}
-\frac{3}{2} \frac{1}{(1-x)_+} -\frac{1+x^2}{1-x} \ln x +\\
& &  +\, 2 + x - \left(\frac{9}{2}+\frac{\pi^2}{3}\right) \delta (1-x)
\Bigg]\\
\delta C_g(x) &=& 2 T_f \left[ (2x-1) \left(\ln \frac{1-x}{x}-1\right)+
2(1-x)\right]
\end{eqnarray}
with $C_F=4/3$ and $T_f=f/2$. Here $f$ denotes, as usual, the number of
active flavors ($f=3$). The NLO expression for the unpolarized
(spin-averaged) structure function $F_1^N(x,Q^2)$ is similar to the one in
(2.2) with $\delta f(x,Q^2)\rightarrow f(x,Q^2)$ and
the unpolarized Wilson coefficients are given, for example,
in \cite{ref13}. Henceforth we shall, as always, use the notation
$\delta q^p\equiv \delta q$ and $q^p\equiv q$. Furthermore the NLO running
coupling is given by
\begin{equation}
\frac{\alpha_s(Q^2)}{4\pi} \simeq \frac{1}{\beta_0 \ln Q^2/\Lambda^2_
{\overline{\rm{MS}}}} -\frac{\beta_1}{\beta_0^3}
\frac{\ln\ln Q^2/\Lambda^2_{\overline{\rm{MS}}}}
{\left(\ln Q^2/\Lambda^2_{\overline{\rm{MS}}}\right)^2}
\end{equation}
with $\beta_0=11-2f/3$, $\beta_1=102-38f/3$ and $\Lambda^{(f)}_
{\overline{\rm{MS}}}$ being given by \cite{ref14}
\begin{displaymath}
\Lambda_{\overline{\rm{MS}}}^{(3,4,5)} = 248, 200, 131\;{\rm{MeV}}\;\;\;.
\end{displaymath}
The number of active flavors $f$ in $\alpha_s(Q^2)$ was fixed by the number
of quarks with $m_q^2\le Q^2$ taking $m_c=1.5\,{\rm{GeV}}$ and
$m_b=4.5\,{\rm{GeV}}$. The marginal charm contribution to $g_1^N$, stemming
from the subprocess $\gamma^* g\rightarrow c\bar{c}$ \cite{ref15}, will
be disregarded throughout. The charm contribution to $F_1^N$ is also small
in the kinematic range covered by present polarization experiments.

For calculating the NLO evolutions of the spin-dependent parton distributions
$\delta f(x,Q^2)$ in (2.2) we have used the well known analytic NLO
solutions in Mellin $n$-moment space (see, e.g., refs.\cite{ref13,ref16,ref17})
with the $n$-th moment being defined by
\begin{equation}
\delta f^n(Q^2) = \int_0^1 dx\; x^{n-1} \delta f(x,Q^2)\;\;\;.
\end{equation}
These $Q^2$-evolutions are governed by the anomalous dimensions\footnote{
Alternatively one can of course use instead the LO and NLO splitting functions
$\delta P_{ij}^{(0)n}=-\delta \gamma_{ij}^{(0)n}/4$ and
$\delta P_{ij}^{(1)n}=-\delta \gamma_{ij}^{(1)n}/8$, respectively
(see, e.g., ref.{\protect{\cite{ref17}}}).}
\begin{eqnarray}
\delta \gamma_{NS}^n &=& \frac{\alpha_s}{4 \pi} \delta \gamma_{qq}^{(0)n}+
\left(\frac{\alpha_s}{4\pi}\right)^2 \delta \gamma_{NS}^{(1)n}(\eta)\;,\;\;
\;\eta = \pm 1\\
\delta \gamma_{ij}^n &=& \frac{\alpha_s}{4 \pi} \delta \gamma_{ij}^{(0)n}+
\left(\frac{\alpha_s}{4\pi}\right)^2 \delta \gamma_{ij}^{(1)n}\;,\;\;
\;i,j = q,g
\end{eqnarray}
whose detailed $n$-dependence will be specified in the Appendix. The
non-singlet (NS) parton densities evolve according to \cite{ref13,ref16}
\begin{equation}
\nonumber
\delta q_{NS\;\eta}^n(Q^2) = \!\Bigg[
1+\frac{\alpha_s(Q^2)-\alpha_s(Q^2_0)} {4\pi}
\left( \frac{\delta \gamma^{(1)n}_{NS}(\eta)}{2 \beta_0}-
\frac{\beta_1\, \delta\gamma_{qq}^{(0)n}}{2\beta_0^2}
\right)\Bigg]
\left( \frac{\alpha_s(Q^2)}{\alpha_s(Q_0^2)}\right)^
{\delta\gamma_{qq}^{(0)n}/
2\beta_0}\!\! \delta q_{NS\;\eta}^n(Q_0^2)
\end{equation}
with the input scale $Q_0^2=\mu^2_{NLO}=0.34\,{\rm{GeV}}^2$ referring to the
radiative \cite{ref14} NLO input ($\mu_{NLO}=\mu_{HO}$) to be discussed
later. Furthermore, opposite to the situation of unpolarized
(spin-averaged) parton distributions \cite{ref16}, $\delta q_{NS\;\eta=+1}$
corresponds to the NS combinations $\delta u -\delta \bar{u}\equiv
\delta u_V$ and $\delta d-\delta \bar{d}\equiv \delta d_V$, while
$\delta q_{NS\;\eta=-1}$ corresponds to the combinations
$\delta q +\delta \bar{q}$ appearing in the NS expressions
\begin{equation}
\delta q_3\equiv (\delta u +\delta \bar{u}) - (\delta d +\delta\bar{d})\;,\;\;
\delta q_8\equiv (\delta u+\delta\bar{u}) +(\delta d+\delta \bar{d})-
2(\delta s+\delta\bar{s})\;\;\;.
\end{equation}
It should be noted that the first ($n=1$) moments $\delta q_{NS-}^1\equiv
\Delta q_{NS-}$ of these latter $SU(3)_f$ diagonal flavor non-singlet
combinations do {\em{not}} renormalize, i.e.\ are independent
of $Q^2$, due to the conservation of the flavor non-singlet axial
vector current, i.e.\ $\delta \gamma_{qq}^{(0)1}=\delta\gamma_{NS}^{(1)1}
(\eta=-1)=0$ (see Appendix). The evolution in the flavor singlet sector, i.e.\
of
\begin{equation}
\delta \Sigma^n(Q^2) \equiv \sum_{q=u,d,s} \left[ \delta q^n(Q^2)+
\delta\bar{q}^n(Q^2)\right]
\end{equation}
and $\delta g^n(Q^2)$ is governed by the anomalous dimension $2\times 2$
matrix (2.9) with the explicit solution given by eq.(2.9) of ref.\cite{ref16}
where $\gamma \rightarrow \delta \gamma$ as given in the Appendix.

Having obtained the analytic NLO solutions for the moments of parton
densities, $\delta f^n(Q^2)$, it is simple to (numerically) Mellin-invert
them to Bj\o rken-$x$ space as described, for example, in \cite{ref16}
or \cite{ref17}. The so obtained $\delta f(x,Q^2)$ have to be convoluted
with the Wilson coefficients in (2.2) to yield the desired $g_1(x,Q^2)$.
Alternatively, one could insert $\delta f^n(Q^2)$ directly into the $n$-th
moment of eq.(2.2),
\setcounter{equation}{1}
\renewcommand{\theequation}{\arabic{section}.\arabic{equation}'}
\begin{eqnarray}
\nonumber
g_1^n(Q^2) &=& \frac{1}{2} \sum_q e_q^2\;
\Bigg\{\delta q^n(Q^2) +\delta \bar{q}^n (Q^2) +\\
&+& \frac{\alpha_s (Q^2)}{2 \pi} \left[ \delta C_q^n\left(
\delta q^n(Q^2)+\delta \bar{q}^n(Q^2)\right) +\frac{1}{f}
\delta C_g^n\;\delta g^n(Q^2)\right]\Bigg\}
\end{eqnarray}
with the moments of (2.4) and (2.5) given by
\setcounter{equation}{3}
\begin{eqnarray}
\!\!\!\!\!\!\delta C_q^n\!\!\!& = &\!\!\!C_F\!
\left[ -S_2(n)+\left(S_1(n)\right)^2\!+\!\left(
\frac{3}{2}-\frac{1}{n(n+1)}\!\right) S_1(n)\!+\frac{1}{n^2}+\frac{1}{2n}+
\frac{1}{n+1}-\frac{9}{2}\right]\\
\delta C_g^n\!\!\!& = &\!\!\!2 T_f
\left[-\frac{n-1}{n(n+1)} \left(S_1(n)+1\right) -\frac{1}{n^2}+
\frac{2}{n(n+1)}\right]
\end{eqnarray}
with $S_k(n)$ defined in the Appendix. The full expression (2.2') can now
be directly (numerically) Mellin-inverted \cite{ref16,ref17} to yield
$g_1(x,Q^2)$ without having to calculate any convolution (2.3).

The LO results are of course entailed in all these expressions given above,
by simply dropping all the obvious higher order terms $(\beta_1,\;
\delta\gamma^{(1)},\; \delta C_{q,g})$ in all relevant equations
stated above.

It should be noted that the first ($n=1$) moment
$\Gamma_1(Q^2)\equiv g_1^1(Q^2)$ in (1.5) is, according to (2.2'), simply given
by
\setcounter{equation}{12}
\renewcommand{\theequation}{\arabic{section}.\arabic{equation}}
\begin{equation}
\Gamma_1(Q^2)=\frac{1}{2} \sum_q e_q^2
\left( 1-\frac{\alpha_s(Q^2)}{\pi}
\right) \left[ \Delta q(Q^2)+\Delta \bar{q}(Q^2)\right]
\end{equation}
where we have used the definition (1.1) and $\delta C_q^1=-3 C_F/2$ and
$\delta C_g^1=0$ according to (2.4') and (2.5'), respectively. Thus, the
total gluon helicity $\Delta g(Q^2)$ does not directly couple to
$\Gamma_1(Q^2)$ due to the vanishing of the integrated gluonic coefficient
function in the $\overline{\rm{MS}}$ factorization scheme. This vanishing
of $\Delta C_g \equiv \delta C_g^1$, which has been some matter of dispute
during the past years (for reviews see, for example, \cite{ref12,ref18,ref19})
originates from the last term in (2.5) proportional to $2(1-x)$. Since this
term derives from the soft non-perturbative collinear region \cite{ref20}
where $k_T^2 \sim m_q^2\ll \Lambda^2$, it has been suggested
[18, 21-23] to absorb it into the definition of the light (non-perturbative)
input (anti)quark distributions
$\delta\!\!\stackrel{(-)}{q}(x,Q^2=Q_0^2)$.
This implies that, instead of $\delta C_g(x)$ in (2.5), one has
\begin{equation}
\delta \tilde{C}_g(x) =2 T_f\,(2x-1) \left(\ln \frac{1-x}{x} -1\right)
\end{equation}
which refers to some different factorization
scheme \cite{ref18,ref22,ref23}
with the $n$-th moment given by
\setcounter{equation}{13}
\renewcommand{\theequation}{\arabic{section}.\arabic{equation}'}
\begin{equation}
\delta\tilde{C}_g^n=\delta C_g^n -2 T_f \frac{2}{n (n+1)}
\end{equation}
and $\delta C_g^n$ given in (2.5'). Thus $\Delta \tilde{C}_g\equiv
\delta \tilde{C}^1_g=-2 T_f$ and $\Delta g(Q^2)$ would couple directly
\cite{ref21} to $\Gamma_1(Q^2)$ in (2.13) via $-(\alpha_s/6\pi)
\Delta g(Q^2)$ according to the gluonic term in the curly brackets of
(2.2') for $f=3$ flavors.
Therefore the gluonic contribution on its own could account for a
reduction \cite{ref18,ref21,ref22} of the Ellis-Jaffe estimate (1.4),
as required by experiment, without the need of a sizeable negative total
(strange) sea helicity as discussed in the Introduction.
One could of course choose to work within this particular
factorization scheme or
any other scheme. In this case, however, one has for
consistency reasons to calculate {\em{all}} polarized NLO
quantities ($\delta C_i^n,\delta \gamma_{ij}^{(1)n}$, etc.),
and not just their first ($n=1$) moments, in
these specific schemes {\em{as well as}}
also NLO subprocesses of purely hadronic reactions to which
the NLO parton distributions are applied to.
The transformation\footnote{
This transformation cannot even be uniquely defined for the spin dependent
case. This is partly in contrast to the unpolarized situation
\protect{\cite{ref13,ref24,ref16}} where the energy-momentum conservation
constraint (for $n=2$) is used together with the assumption of its
analyticity in $n$.}
$\delta C_g^n \rightarrow \delta \tilde{C}_g^n$ in (2.14') implies of
course also a
corresponding modification \cite{ref13,ref24} of the NLO anomalous
dimensions $\delta \gamma_{ij}^{(1)n}\rightarrow
\delta \tilde{\gamma}_{ij}^{(1)n}$.
\section{Quantitative NLO Analysis}
\renewcommand{\theequation}{\arabic{section}.\arabic{equation}}
\setcounter{equation}{0}
In fixing the polarized NLO input parton distributions
$\delta f(x,Q^2=\mu^2_{NLO})$ we follow closely our recent LO analysis
\cite{ref1}. We still prefer to work with the directly measured asymmetry
$A_1^N(x,Q^2)$ in (2.1), rather than with the derived $g_1^N(x,Q^2)$,
since possible non-perturbative (higher twist) contributions are
expected to partly cancel in the ratio of structure functions
appearing in $A_1^N(x,Q^2)$, in contrast to the situation for $g_1^N(x,Q^2)$.
Therefore we shall use all presently available data [4-9] in the small-$x$
region where
$Q^2\,$\raisebox{-1mm}{${\stackrel{\textstyle >}{\sim}}$}$\,1\,{\rm{GeV}}^2$
without bothering about lower cuts
in $Q^2$ usually introduced in order to avoid possible higher twist effects as
mandatory for analyzing $g_1^N(x,Q^2)$ in the low-$Q^2$ region. The
analysis affords some well established set of unpolarized NLO parton
distributions $f(x,Q^2)$ for calculating $F_1^N(x,Q^2)$ in (2.1) which will
be adopted from ref.\cite{ref14}, i.e.\ our recent updated
NLO ($\overline{\rm{MS}}$) dynamical distributions valid down to the radiative
input scale $Q^2=\mu^2_{NLO}=0.34\,{\rm{GeV}}^2$.

The searched for polarized NLO (as well as LO)
parton distributions $\delta f(x,Q^2)$,
compatible with present data [4-9] on $A_1^N(x,Q^2)$, are constrained by the
positivity requirements (1.3) and for the $SU(3)_f$ symmetric
'standard' scenario by the sum rules
\begin{eqnarray}
&&\Delta u + \Delta \bar{u} - \Delta d - \Delta \bar{d} =
g_A = F + D = 1.2573 \pm 0.0028 \\
&&\Delta u + \Delta \bar{u} + \Delta d + \Delta \bar{d} -
2(\Delta s + \Delta \bar{s}) = 3F -D = 0.579 \pm 0.025
\end{eqnarray}
with the first moment $\Delta f$ defined in (1.1)
and the values of $g_A$ and $3F-D$ taken from \cite{ref25}.
It should be remembered that the first moments $\Delta q_{3,8}$ of the flavor
non-singlet combinations (2.11) which appear in (3.1) and (3.2) are
$Q^2$-independent also in NLO due to $\delta \gamma_{NS}^{(1)1}(\eta =-1)=0$,
according to eq.(A.9).

As a plausible alternative to the full $SU(3)_f$ symmetry between charged
weak and neutral axial currents required for deriving the 'standard'
constraints (3.1) and (3.2), we consider a 'valence' scenario \cite{ref1,ref2}
where this flavor symmetry is broken and which is based on the assumption
\cite{ref2} that the flavor changing
hyperon $\beta$-decay data fix only the total helicity of
{\em{valence}} quarks
$\Delta q_V(Q^2) \equiv \Delta q - \Delta \bar{q}$ :
\setcounter{equation}{0}
\renewcommand{\theequation}{\arabic{section}.\arabic{equation}'}
\begin{eqnarray}
\Delta u_V(\mu^2_{NLO})-
\Delta d_V(\mu^2_{NLO}) &=& g_A = F+D= 1.2573 \pm 0.0028 \\
\Delta u_V(\mu^2_{NLO})+\Delta d_V(\mu^2_{NLO})
&=& 3F-D  = 0.579 \pm 0.025\;\;\;.
\end{eqnarray}
Although at the input scale $\Delta \bar{u} (\mu^2_{NLO})=
\Delta\bar{d} (\mu^2_{NLO})$, isospin symmetry will be
(marginally) broken by the NLO evolution, i.e.\ $\Delta \bar{u}(Q^2)\neq
\Delta\bar{d}(Q^2)$ for $Q^2>\mu^2_{NLO}$.
In addition we shall assume a maximally $SU(3)_f$ broken polarized
strange sea input $\delta s(x,\mu^2_{NLO}) =\delta\bar{s}(x,\mu^2_{NLO})=0$
in our 'valence' scenario, which in addition is compatible with the
$SU(3)_f$ broken unpolarized radiative input $s(x,\mu^2_{NLO})=0$ of
ref.\cite{ref14}. Such a choice is feasible in the 'valence' scenario since,
due to eq.(2.13) and (2.11), we have in general
\setcounter{equation}{2}
\renewcommand{\theequation}{\arabic{section}.\arabic{equation}}
\begin{equation}
\Gamma_1^{p,n}(Q^2) = \left[ \pm \frac{1}{12} \Delta q_3 +\frac{1}{36}
\Delta q_8 +\frac{1}{9} \Delta \Sigma(Q^2) \right]
\left(1-\frac{\alpha_s(Q^2)}{\pi}\right)\;\;\;.
\end{equation}
Since $\Delta u_V(Q^2) -\Delta d_V(Q^2)$ and $\Delta u_V(Q^2)+
\Delta d_V(Q^2)$ decrease only marginally with $Q^2$ as compared to their
input values in (3.1') and (3.2') [cf.\ Table 1' below], and furthermore
the dynamical isospin breaking $\Delta \bar{u} \neq \Delta \bar{d}$
at $Q^2>\mu_{NLO}^2$ is small, eq.(3.3) can be approximated by
\begin{eqnarray}
\nonumber
\Gamma_1^{p,n} (Q^2) & \simeq & \Bigg[ \pm \frac{1}{12} (F+D) +
\frac{5}{36} (3F -D) +\\
&& + \frac{1}{18} \left( 10 \Delta \bar{q} (Q^2) +\Delta s(Q^2)+
\Delta\bar{s} (Q^2)\right) \Bigg] \left( 1-\frac{\alpha_s(Q^2)}{\pi}
\right)\;\;\;\;.
\end{eqnarray}
Therefore a light polarized sea $\Delta \bar{q} \equiv
(\Delta \bar{u} +\Delta \bar{d})/2 <0$
can account for the reduction of the Ellis-Jaffe estimate (1.4)
for $\Gamma_1^p(Q^2)$, say, as required by recent experiments \cite{ref5,ref6}.

In the 'standard' scenario we need, on the contrary, a finite sizeable
$\Delta s(Q^2)<0$ since, due to eq.(2.13) and (2.11),
\begin{equation}
\Gamma_1^{p,n}(Q^2)=\left[\pm \frac{1}{12} \Delta q_3 +\frac{5}{36}
\Delta q_8 +\frac{1}{3} \left(\Delta s(Q^2)+\Delta \bar{s}(Q^2)
\right)\right] \left(1-\frac{\alpha_s(Q^2)}{\pi}\right)
\end{equation}
with the $Q^2$-independent flavor non-singlet combinations $\Delta q_{3,8}$
being entirely fixed by eqs.(3.1) and (3.2); for the singlet combination in
(1.2) we used $\Delta\Sigma (Q^2)\;=\;\Delta q_8\;+ $ \linebreak
$3\left(\Delta s(Q^2)+
\Delta \bar{s}(Q^2)\right)$. It should be noted that, in contrast to the
LO \cite{ref1} case, a finite $\Delta s(Q^2)$ will be generated
dynamically in NLO for $Q^2>\mu^2_{NLO}$ even for a vanishing input
$\Delta s(\mu^2_{NLO})=0$ due to the non-vanishing NLO $\Delta
\gamma_{qq}^{(1)}$ in (A.9): The resulting 'dynamical' $\Delta s(Q^2)<0$
is about an order of magnitude too small to comply with recent
experiments \cite{ref5,ref6} which typically yield, for example,
$\Gamma_1^p(Q^2=3\,{\rm{GeV}}^2)\simeq 0.12-0.13$, i.e.\ sizeably smaller
than the naive estimate (1.4). We therefore have to implement a
{\em{finite}} strange sea input $\Delta s(\mu^2_{NLO})=
\Delta\bar{s}(\mu^2_{NLO})<0$ for the 'standard' scenario, in order to
arrive at $\Delta s(Q^2=3-10\,{\rm{GeV}}^2) \simeq -0.05$ as required
\cite{ref1} by recent experiments.

Apart from applying the above scenarios for the polarized input distributions
to $A_1^N(x,Q^2)$ rather than to $g_1^N(x,Q^2)$, the main ingredient of
our NLO analysis is the implementation of the positivity constraints
(1.3) down to \cite{ref14} $Q^2=\mu^2_{NLO}=0.34\,{\rm{GeV}}^2$
(and to $Q^2=\mu^2_{LO}=0.23\,{\rm{GeV}}^2$ in LO) which is not
guaranteed in the usual (LO) studies done so far (recently, e.g., in
[26-29]) restricted to $Q^2\ge Q_0^2=1-4\,{\rm{GeV}}^2$. We follow here
the radiative (dynamical) concept which resulted in the successful
small-$x$ predictions of unpolarized parton distributions as measured
at HERA \cite{ref14,ref16,ref30}. A further advantage of this analysis
is the possibility to study the $Q^2$-dependence of $A_1^N(x,Q^2)$ in the
small-$x$ region over a wide range of $Q^2$ \cite{ref6}
which might be also relevant for
the forthcoming polarized experiments at HERA. In addition it will be
important to learn about the reliability of perturbative calculations by
comparing the LO with the NLO results;
a reasonable perturbative stability of all
radiative model predictions will be indeed observed for
{\em{measurable}} quantities such as $A_1^N(x,Q^2)$ and
$g_1^N(x,Q^2)$, as is the case for spin-averaged deep inelastic structure
functions \cite{ref14,ref30}.

Turning to the determination of the polarized NLO (LO) parton distributions
$\delta f(x,Q^2)$ it is helpful to consider some reasonable theoretical
constraints concerning the sea and gluon distributions, in particular in
the relevant small-$x$ region where only rather scarce data exist at
present, such as color coherence of gluon couplings at $x\simeq 0$ and
helicity retention properties of valence densities as $x\rightarrow 1$
\cite{ref31}. We follow here very closely the procedure and ans\"{a}tze
of ref.\cite{ref1}.
Subject to these constraints we employ the following general ansatz for the
NLO (LO) polarized parton distributions:
\begin{eqnarray}
\delta q_V(x,\mu^2)&=& N_{q_V}\,x^{a_{q_V}}\,q_V(x,\mu^2) \nonumber \\
\delta \bar{q}(x,\mu^2) &=& N_{\bar{q}}\,x^{a_{\bar{q}}}
\,(1-x)^{b_{\bar{q}}}\,\bar{q}(x,\mu^2) \nonumber \\
\delta s(x,\mu^2)&=&\delta \bar{s}(x,\mu^2) = N_s\,\delta \bar{q}(x,\mu^2)
\nonumber \\
\delta g(x,\mu^2) &=& N_g\, x^{a_g}\,(1-x)^{b_g}\, g(x,\mu^2)
\end{eqnarray}
with the NLO (LO) unpolarized input densities being taken from ref.\cite{ref14}
and, for obvious reasons, we have not taken into account any $SU(2)_f$
breaking input ($\delta \bar{u} \neq \delta \bar{d}$) as is apparent from
our ansatz for $\delta \bar{q}\equiv \delta\bar{u} \equiv \delta \bar{d}$
proportional to $\bar{q}\equiv(\bar{u}+\bar{d})/2$ which should be
considered as the reference light sea distribution for the positivity
requirement (1.3).
In the `standard' scenario our optimal NLO densities at
$Q^2=\mu^2_{NLO}=0.34\,{\rm{GeV}}^2$ are given by
\begin{eqnarray}
\nonumber N_{u_V} &=& 0.6586\;\;,
\;\;a_{u_V} = 0.18\;\;;\;\;N_{d_V}=-0.3392\;\;,
\;\;a_{d_V}=0\\
\nonumber N_{\bar{q}} &=& -0.525\;\;,\;\;a_{\bar{q}} = 0.51\;\;,\;\;
b_{\bar{q}}=0\;\;;\;\;N_s=1\\
N_g &=& 10.47\;\;,\;\; a_g = 1.1\;\;,\;\;b_g = 4.3
\end{eqnarray}
\noindent
corresponding to $\chi^2=104.5/125$ d.o.f.\ and respecting eqs.(3.1) and (3.2)
which are the basis of almost all analyses performed so far.$^{3,\,4}$
\stepcounter{footnote}\footnotetext{
It is interesting to note that, within our radiative approach with its
longer $Q^2$-evolution 'distance', a {\em{finite}} (negative)
strange sea input $\delta s(x,\mu^2_{NLO})$ is
{\em{always}} required by
present data even if one uses $\delta \tilde{C}_g$ in (2.14) or (2.14')
[21-23].
For the latter case, $\delta s$ has to be at least half as large as in (3.5)
which is based on the $\overline{\rm{MS}}$ $\delta C_g$ in (2.5) or (2.5').
This holds true even for a maximally saturated input gluon
$[\delta g(x,\mu^2_{NLO})=g(x,\mu^2_{NLO})]$ to be discussed below.}
\stepcounter{footnote}\footnotetext{
It should be noted that our fit result $N_s=1$ in (3.7) implies via (3.6) an
$SU(3)_f$ symmetric sea input. In NLO at $Q^2>\mu^2_{NLO}$ this symmetry
becomes dynamically broken via the $Q^2$ - evolution. In view of present
scarce data such $SU(3)_f$ [and $SU(2)_f$] breaking effects are entirely
negligible for quantitative analyses.}
Since new polarized $p$ and $d$ data have appeared very recently
\cite{ref6,ref7}, we have also again determined the LO distributions in
the 'standard' scenario at $Q^2=\mu^2_{LO}=0.23\,{\rm{GeV}}^2$:
\setcounter{equation}{6}
\renewcommand{\theequation}{\arabic{section}.\arabic{equation}'}
\begin{eqnarray}
\nonumber N_{u_V} &=& 0.6392\;\;,
\;\;a_{u_V} = 0.16\;\;;\;\;N_{d_V}=-0.3392\;\;,
\;\;a_{d_V}=0\\
\nonumber N_{\bar{q}} &=& -1.348\;\;,\;\;a_{\bar{q}} = 0.93\;\;,\;\;
b_{\bar{q}}=0.09\;\;;\;\;N_s=1\\
N_g &=& 11.8\;\;,\;\; a_g = 0.76\;\;,\;\;b_g = 6.84
\end{eqnarray}
\setcounter{equation}{7}%
\renewcommand{\theequation}{\arabic{section}.\arabic{equation}}%
\noindent
corresponding to $\chi^2=108.0/125$ d.o.f.\ and which supercedes our LO
input fit of ref.\cite{ref1}.
The fact that $\delta s(x,\mu^2)\neq 0$ in
(3.7) and (3.7'), i.e.\ $N_s=1$, contradicts somewhat our purely
radiative input \cite{ref14}
$s(x,\mu^2)=\bar{s}(x,\mu^2)=0$, but for
$Q^2\,$\raisebox{-1mm}{${\stackrel{\textstyle >}{\sim}}$}$\,
1\,{\rm{GeV}}^2$
the positivity inequality (1.3) is already satisfied,
in particular at large values of $x$. In this respect the input for the
'valence' scenario with the extreme $SU(3)_f$ breaking ansatz
$\delta s(x,\mu^2)=0$ is more consistent as far as our radiative
(dynamical) approach is concerned:
For the $SU(3)_f$ broken 'valence' scenario, based on the
constraints (3.1') and (3.2'), our optimal NLO input corresponds to the
following parameters in (3.6) at $\mu^2_{NLO}=0.34\,{\rm{GeV}}^2$:
\begin{eqnarray}
\nonumber N_{u_V} &=& 0.6708\;\;,
\;\;a_{u_V} = 0.19\;\;;\;\;N_{d_V}=-0.3693\;\;,
\;\;a_{d_V}=0.03\\
\nonumber N_{\bar{q}} &=& -0.6\;\;,\;\;a_{\bar{q}} = 0.5\;\;,\;\;
b_{\bar{q}}=0\;\;;\;\;N_s=0\\
N_g &=& 9.87\;\;,\;\; a_g = 1.05\;\;,\;\;b_g = 4.44
\end{eqnarray}
\noindent
corresponding to $\chi^2=104.6/125$ d.o.f. Similarly our new LO input at
$\mu^2_{LO}=0.23\,{\rm{GeV}}^2$ in the 'valence' scenario is given by
\setcounter{equation}{7}
\renewcommand{\theequation}{\arabic{section}.\arabic{equation}'}
\begin{eqnarray}
\nonumber N_{u_V} &=& 0.6272\;\;,
\;\;a_{u_V} = 0.15\;\;;\;\;N_{d_V}=-0.3392\;\;,
\;\;a_{d_V}=0\\
\nonumber N_{\bar{q}} &=& -1.455\;\;,\;\;a_{\bar{q}} = 0.88\;\;,\;\;
b_{\bar{q}}=0.13\;\;;\;\;N_s=0\\
N_g &=& 7.4\;\;,\;\; a_g = 0.6\;\;,\;\;b_g = 5.93
\end{eqnarray}
\setcounter{equation}{8}%
\renewcommand{\theequation}{\arabic{section}.\arabic{equation}}%
\noindent
corresponding to $\chi^2=108.6/125$ d.o.f.\ and which supercedes our LO
input fit of ref.\cite{ref1}.
Finally, similarly agreeable NLO
'valence' scenario fits to all present asymmetry data shown below (with
a total $\chi^2$ of 106.2 to 107.6 for 125 data points)
can be also obtained for
a fully saturated (inequality (1.3)) gluon input
$\delta g(x,\mu_{NLO}^2)=g(x,\mu_{NLO}^2)$ as well as for the less saturated
$\delta g(x,\mu_{NLO}^2)=xg(x,\mu_{NLO}^2)$. A purely dynamical \cite{ref32}
input $\delta g(x,\mu_{NLO}^2)=0$ is also compatible with present
data, but such a choice seems to be unlikely in view of $\delta
\bar{q}(x,\mu_{NLO}^2)\neq 0$; it furthermore results in an unphysically
steep \cite{ref32} $\delta g(x,Q^2 > \mu_{NLO}^2)$, being mainly
concentrated in the very small-$x$ region $x<0.01$, as in the
corresponding case \cite{ref16,ref33} for the unpolarized parton distributions
in disagreement with experiment.
Similar remarks hold also for a LO analysis as well as
for the 'standard' scenario.

A comparison of our results with the data on $A_1^N(x,Q^2)$ is presented in
Fig.1. The LO and NLO results in the 'standard' scenario are perturbatively
stable and almost indistinguishable. The same holds true for the results
in the 'valence' scenario which are not shown separately since they almost
coincide with the 'standard' ones in Fig.1.
As already mentioned,
fit results using a 'saturated' gluon $\delta g=g$ or $\delta g = xg$,
or even $\delta g=0$
at $Q^2=\mu_{LO,NLO}^2$ are very similar to the ones shown in Fig.1.
Note that $A_1^N(x,Q^2)\rightarrow {\rm{const.}}$ as $x\rightarrow 1$.
The recently measured $Q^2$ - dependence of $A_1^N(x,Q^2)$ \cite{ref6} is
compared with our theoretical results down to $Q^2=0.4\,{\rm{GeV}}^2$ in
Fig.2. The difference between our LO and NLO results in the small-$Q^2$
region is mainly due to different LO ($\mu^2_{LO}=0.23\,{\rm{GeV}}^2)$
and NLO $(\mu^2_{NLO}=0.34\,{\rm{GeV}}^2)$ input scales.
The more detailed
$Q^2$-dependence of $A_1^N(x,Q^2)$ is presented in Fig.3 for some typical
fixed $x$ values for
$1\,$\raisebox{-1mm}{${\stackrel{\textstyle <}{\sim}}$}$\,Q^2 \le 20
{\rm{GeV}}^2$ relevant for present
experiments. The predicted scale-violating NLO $Q^2$-dependence is
similar to the LO one; for $x>0.01$ this is also the case
for the two rather different input scenarios (3.1), (3.2) and
(3.1'), (3.2'). In the $(x,Q^2)$ region of present data
[4-9], $A_1^p(x,Q^2)$ increases with
$Q^2$ for $x>0.01$. Therefore, since most present data in the small-$x$
region correspond to small values of $Q^2\,
$\raisebox{-1mm}{${\stackrel{\textstyle >}{\sim}}$}$\,0.5$\,GeV$^2$,
the determination of $g_1^p(x,Q^2)$ at a larger fixed
$Q^2$ (5 or 10 GeV$^2$, say) by assuming $A_1^p(x,Q^2)$ to be independent
of $Q^2$, as is commonly done [4-9]
(except for the recent last reference of \cite{ref6}),
is misleading and might lead to
an {\em{under}}estimate of $g_1^p$ by as much as about $20\%$,
in particular in the small-$x$ region
$x\,$\raisebox{-1mm}{${\stackrel{\textstyle >}{\sim}}$}$\, 0.02$.
(For $x\,$\raisebox{-1mm}{${\stackrel{\textstyle <}{\sim}}$}$\, 0.01$
the effect will be opposite.)
The situation is opposite, although less pronounced, for $-A^n_1(x,Q^2)$
shown in Fig.3. This implies that $\left| g_1^n(x,Q^2)\right|$ might be
{\em{over}}estimated at larger fixed $Q^2$ by assuming $A_1^n(x,Q^2)$,
as measured at small $Q^2$, to be independent of $Q^2$. It is obvious that
the assumption of approximate scaling for $A_1(x,Q^2)$ is therefore
unwarranted and, in any case, theoretically not justified
as soon as gluon and sea densities become relevant, due to the very different
polarized and unpolarized splitting functions (anomalous dimensions)
in the flavor singlet sector.

In Fig.4 we compare our NLO results for $g_1^N(x,Q^2)$ with EMC, SMC
and SLAC-E142/E143 data as well as with our
LO results which are similar to our original LO results \cite{ref1}.
The reason why the LO results are partly larger by more than about $10\%$
than the NLO ones is mainly due to the LO approximation where
$R^N=0$ in (2.1). Although the agreement between the NLO results and experiment
has been significantly improved, the EMC \cite{ref4} and E143 \cite{ref6}
'data' at fixed values of $Q^2$ fall still below our NLO predictions in the
small-$x$ region. This is partly due to the fact that the original
small-$x$ $A_1^p$-data at small $Q^2$ have been extrapolated \cite{ref4,ref6}
to a larger fixed value of $Q^2$ by
assuming $A_1^p (x,Q^2)$ to be independent of $Q^2$. According to
the increase of $A_1^p$ with $Q^2$ in Fig.3, such an assumption underestimates
$g_1^p$ in the small-$x$ region at larger $Q^2$. On the contrary,
our results for $g_1^{p,d}$ do not show such a disagreement in the
small-$x$ region when compared with the SMC data \cite{ref5,ref7} in
Figs.\ 4a and 4b
where each data point corresponds to a different value of $Q^2$ since
no attempt has been made to extrapolate $g_1^N(x,Q^2)$ to a fixed
$Q^2$ from the originally measured $A_1^N(x,Q^2)$.
Our predictions for the NLO parton distributions at
the input scale $Q^2=\mu_{NLO}^2$ in eq.(3.6)
with the fit parameters given in (3.7) and (3.8) are shown in Fig.5a;
the corresponding LO inputs at $Q^2=\mu^2_{LO}$ in eq.(3.6) with the
fit parameters given in (3.7') and (3.8') are shown in Fig.5b.
The polarized input densities in Figs.\ 5a and 5b are compared with our
reference unpolarized NLO and LO dynamical input densities of ref.\cite{ref14}
which satisfy of course the positivity requirement (1.3) as is obvious from
eq.(3.6). The distributions at $Q^2=4\,{\rm{GeV}}^2$, as obtained from these
inputs at $Q^2=\mu^2$ for the two scenarios considered, are shown in Fig.6.
Since not even the polarized NLO gluon density
$\delta g(x,Q^2)$ is strongly constrained by present experiments,
we compare our gluons at $Q^2=4$ GeV$^2$ in Fig.7
with the ones which originate
from imposing extreme inputs at $Q_0^2=\mu_{NLO}^2$, such as
$\delta g=g$, $\delta g=xg$ and $\delta g=0$, instead of the one
in (3.6) for the 'valence' scenario.
The results are similar if these extreme gluon-inputs
are taken for the `standard' scenario in (3.6), and the variation
of $\delta g(x,Q^2)$ allowed by present experiments is indeed
sizeable.
This implies, in particular, that the $Q^2$-evolution of $g_1(x,Q^2)$
below the experimentally accessible $x$-range is {\em{not}} predictable
for the time being.

Finally let us turn to the first moments (total polarizations)
$\Delta f(Q^2)$ of our polarized parton distributions, as
defined in (1.1), and the resulting $\Gamma_1^{p,n}(Q^2)$ in (3.3).
It should be recalled that, in contrast to the LO, the first moments of the
NLO (anti)quark densities do renormalize, i.e.\ are $Q^2$-dependent, due
to the non-vanishing of the 2-loop $\delta\gamma_{qq}^{(1)1}$ in
(A.9) and $\delta \gamma_{NS}^{(1)1}(\eta=+1)$ in (2.10). Let us discuss
the two scenarios in turn:
\begin{description}
\item[`standard' scenario:] From the input distributions (3.6)
together with (3.7), being
constrained by (3.1) and (3.2), one infers in NLO
\begin{center}
\renewcommand{\arraystretch}{1.5}
\begin{tabular}{c||c|c|c|c|c|c||c|c}
$Q^2\,({\rm{GeV}}^2)$ & $\Delta u_V$ & $\Delta d_V$ & $\Delta \bar{q}$ &
$\Delta s= \Delta \bar{s}$ & $\Delta g$ & $\Delta \Sigma $&
$\Gamma_1^p$ & $\Gamma_1^n$ \\\hline
$\mu^2_{NLO}$ & 0.9181 & -0.3392 & -0.0660 & -0.0660 & 0.507 &
0.183 & 0.1136 & -0.0550\\
1 & 0.915 & -0.338 & -0.067 & -0.068 & 0.961 & 0.173 & 0.124 &
-0.061 \\
4 & 0.914 & -0.338 & -0.068 & -0.068 & 1.443 & 0.168 & 0.128 &
-0.064 \\
10& 0.914 & -0.338 & -0.068 & -0.069 & 1.737 & 0.166 & 0.130 &
-0.065 \\
\end{tabular}
\end{center}
Table 1.\ First moments $\Delta f$ of polarized NLO parton densities
$\delta f(x,Q^2)$ and of $g_1^{p,n}(x,Q^2)$ as predicted in the
'standard' scenario. Note that the marginal differences for $\Delta\bar{q}$
and $\Delta s$ indicate the typical amount of dynamical $SU(3)_f$
breaking as mentioned in footnote 4.

\noindent
In LO the input distributions (3.6) together with (3.7') imply
\begin{eqnarray}
\nonumber
\Delta u_V=0.9181\;\;,\;\;\Delta d_V=-0.3392\;\;,\;\;,\Delta\bar{q}=
\Delta s=\Delta \bar{s}=-0.0587\;\;,\\
\Delta g(\mu^2_{LO})=0.362\;\;,\;\;\Delta g(4\,{\rm{GeV}}^2)=1.273\;\;,\;\;
\Delta g(10\,{\rm{GeV}}^2)=1.570
\end{eqnarray}
which result in $\Delta \Sigma=0.227$. This gives, using eq.(3.5) without
the factor $\alpha_s/\pi$ together with (3.1) and (3.2),
\begin{equation}
\Gamma_1^p=0.1461\;\;,\;\;\Gamma_1^n=-0.0635
\end{equation}
which is similar to our previous LO result \cite{ref1}. Both our NLO results
in Table 1 and the LO ones in (3.10) are
in satisfactory agreement with recent SMC
and EMC measurements \cite{ref4, ref5,
ref7}
\setcounter{equation}{10}
\renewcommand{\theequation}{\arabic{section}.\arabic{equation}}
\begin{equation}
\Gamma_1^p(10\,{\rm{GeV}}^2) = 0.142\pm 0.008 \pm 0.011, \:\:\:
\Gamma_1^n(10\,{\rm{GeV}}^2) =  -0.063 \pm 0.024 \pm 0.013
\end{equation}
as well as with the most recent E143 data \cite{ref9} implying
$\Gamma_1^n(2\,{\rm{GeV}}^2)=-0.037 \pm 0.008 \pm 0.011$.
\item[`valence' scenario:] From the input distributions (3.6)
together with (3.8), being
constrained by (3.1') and (3.2'), one infers in NLO
\begin{center}
\renewcommand{\arraystretch}{1.5}
\begin{tabular}{c||c|c|c|c|c|c||c|c}
$Q^2\,({\rm{GeV}}^2)$ & $\Delta u_V$ & $\Delta d_V$ & $\Delta \bar{q}$ &
$\Delta s= \Delta \bar{s}$ & $\Delta g$ & $\Delta \Sigma $&
$\Gamma_1^p$ & $\Gamma_1^n$ \\\hline
$\mu^2_{NLO}$ & 0.9181 & -0.3392 & -0.0778 & 0 & 0.496 & 0.268 &
0.1142 & -0.0544\\
1 & 0.915 & -0.338 & -0.080 &$-2.5\times 10^{-3}$&
0.982 & 0.252 & 0.124 & -0.061 \\
4 & 0.914 & -0.338 & -0.081 &$-3.5\times 10^{-3}$&
1.494 & 0.245 & 0.128 & -0.064 \\
10& 0.914 & -0.338 & -0.081 &$-3.8\times 10^{-3}$&
1.807 & 0.244 & 0.130 & -0.065 \\
\end{tabular}
\end{center}
Table 1'.\ First moments $\Delta f$ of polarized NLO parton densities
$\delta f(x,Q^2)$ and of $g_1^{p,n}(x,Q^2)$ as predicted in the
'valence' scenario.

\noindent
In LO the input distributions (3.6) together with (3.8') imply
\setcounter{equation}{8}
\renewcommand{\theequation}{\arabic{section}.\arabic{equation}'}
\begin{eqnarray}
\nonumber
\Delta u_V=0.9181\;\;,\;\;\Delta d_V=-0.3392\;\;,\;\;,\Delta\bar{q}=
-0.0712\;\;,\;\;\Delta s=\Delta \bar{s}=0\;\;,\\
\Delta g(\mu^2_{LO})=0.372\;\;,\;\;\Delta g(4\,{\rm{GeV}}^2)=1.361\;\;,\;\;
\Delta g(10\,{\rm{GeV}}^2)=1.684
\end{eqnarray}
which result in $\Delta \Sigma=0.294$. This gives, using eq.(3.4) without
the factor $\alpha_s/\pi$ together with (3.1') and (3.2'),
\begin{equation}
\Gamma_1^p=0.1456\;\;,\;\;\Gamma_1^n=-0.0639
\end{equation}
which is again similar to our previous LO result \cite{ref1}.
Both our NLO results in Table 1' and the LO ones in (3.10')
compare again well with the experimental results in (3.11).
\end{description}
Apart from the $Q^2$-dependent $\Delta g(Q^2)$ in LO and NLO, the
$Q^2$-dependent first moments of NLO (anti)quark densities in Table 1 and
1' should be compared with the $Q^2$-independent LO results
as discussed in the Introduction
which, in absolute magnitude, are similar to the NLO ones.
Although present scarce data obviously cannot uniquely fix the polarized
sea and gluon densities, our optimal fits favor a sizeable total gluon
helicity, $\Delta g(10\,{\rm{GeV}}^2)\simeq 1.7$, despite the
fact that $\Delta g(Q^2)$ decouples from the full first moment
$\Gamma_1(Q^2)$ in (2.13) in the $\overline{\rm{MS}}$ scheme.

In both scenarios the Bj\o rken sum rule manifestly holds
in LO due to our constraints (3.1) and (3.1'), 
and together with the NLO correction due to
eq.(3.3) we have
\setcounter{equation}{11}
\renewcommand{\theequation}{\arabic{section}.\arabic{equation}}
\begin{equation}
\Gamma_1^p(Q^2) - \Gamma_1^n(Q^2) =\frac{1}{6} g_A \left(
1- \frac{\alpha_s(Q^2)}{\pi}\right)\;\;\;.
\end{equation}
It is also interesting to observe that at our low input scales
$Q^2=\mu^2_{LO,NLO}=0.23,\,0.34\,{\rm{GeV}}^2$
the nucleon's spin is dominantly carried just by the
total helicities of quarks and gluons, $\frac{1}{2}
\Delta \Sigma (\mu^2_{NLO})+ \Delta g(\mu^2_{NLO})
\simeq 0.6$ according to Tables 1 and 1',
and $\frac{1}{2} \Delta \Sigma+\Delta g(\mu^2_{LO}) \simeq 0.5$
according to eqs.(3.9) and (3.9'),
which implies for the helicity sum rule
\begin{equation}
\frac{1}{2} = \frac{1}{2} \Delta \Sigma (Q^2) + \Delta g(Q^2) +
L_z(Q^2)
\end{equation}
that $L_z(\mu^2_{LO,NLO})\simeq 0$. The approximate vanishing of the latter
non-perturbative angular momentum, being build up from the intrinsic $k_T$
carried by partons, is intuitively expected for low (bound-state-like)
scales but not for $Q^2\gg \mu^2_{LO,NLO}$.
\section{Summary}
Based on a recent complete NLO calculation \cite{ref11}
of all spin-dependent two-loop
splitting functions $\delta P_{ij}^{(1)}(x)$, $i,j=q,g$ (or, equivalently,
anomalous dimensions $\delta \gamma_{ij}^{(1)}$) in the conventional
$\overline{\rm{MS}}$ factorization scheme, we have first presented a
consistent NLO formulation of the $Q^2$-evolution of polarized parton
distributions. For calculational purposes we have concentrated on
(Mellin) $n$-moments of structure functions where the solutions
of the NLO evolution equations can be obtained analytically for the parton
densities. Using these formal results we have performed a quantitative NLO
analysis of the longitudinal spin asymmetry $A_1^{p,n}(x,Q^2)$ and of
$g_1^{p,n}(x,Q^2)$, and we have updated our previous LO results \cite{ref1}.
Within the whole relevant $x$- and $Q^2$-region
($x\;$\raisebox{-1mm}{${\stackrel{\textstyle >}{\sim}}$}$\,
10^{-3}$,
$Q^2\,$\raisebox{-1mm}{${\stackrel{\textstyle >}{\sim}}$}$\,
1\,{\rm{GeV}}^2$) we found a remarkable
perturbative stability between LO and NLO results. The scale violating
$Q^2$-dependence of $A_1^{p,n}(x,Q^2)$ turned out to be similar to
the one obtained in LO and is {\em{non}}-negligible
for $(x,Q^2)$ values relevant for present data. The assumption of
approximate scaling for $A_1(x,Q^2)$ is therefore unwarranted and
theoretically not justified. We presented two plausible sets of
polarized LO and NLO
($\overline{\rm{MS}}$) parton densities $\delta f(x,Q^2)$
which describe all presently available data very well. In contrast to
polarized quark and antiquark densities, the gluon density $\delta g(x,Q^2)$
is rather weakly constrained by present data.
Our optimal fits, however, favor a rather sizeable total gluon helicity,
e.g., $\Delta g(Q^2=10\,{\rm{GeV}}^2)\simeq 1.7$.
It should be reemphasized that only processes where $\delta g$ occurs
{\em{directly}} already in LO (with no $\delta q$ and $\delta
\bar{q}$ contributions present) appear to be the most promising sources
for measuring $\delta g(x,Q^2)$. This is the case for
$\gamma^* (\gamma) \delta g \rightarrow c\bar{c}$ responsible for open
charm or $J/\Psi$ production (see, e.g., ref.\cite{ref15}).
Our results demonstrate the compatibility of our restrictive radiative
model, cf.\ eq.(1.3), down to $Q^2=\mu^2_{NLO}=0.34\,{\rm{GeV}}^2$
and to $Q^2=\mu^2_{LO}=0.23\,{\rm{GeV}}^2$, with
present measurements of deep inelastic spin asymmetries and structure
functions.

A {\sc{Fortran}} package containing our optimally fitted 'standard'
and 'valence' NLO ($\overline{\rm{MS}}$) as well as LO
distributions can be obtained
by electronic mail from \linebreak
stratmann@het.physik.uni-dortmund.de or
vogelsang@v2.rl.ac.uk.
\section*{Acknowledgements}
We are grateful to Willy van Neerven for helpful discussions about the
complete NLO($\overline{\rm{MS}}$) calculation of the spin-dependent
splitting functions. This work has been supported in part by the
'Bundesministerium f\"{u}r Bildung, Wissenschaft, Forschung und
Technologie', Bonn.
\section*{Appendix}
\setcounter{equation}{0}
\renewcommand{\theequation}{\rm{A}.\arabic{equation}}
The spin-dependent LO anomalous dimensions (splitting ${\rm{functions}}^1$)
have been originally calculated in \cite{ref34,ref35} and are given by
\begin{eqnarray}
\nonumber
\delta\gamma^{(0)n}_{qq} &=& 4 C_F \left[ 2 S_1(n) -\frac{1}{n(n+1)}-
\frac{3}{2}\right]\\
\nonumber
\delta\gamma_{qg}^{(0)n} & = & - 8 T_f\, \frac{n-1}{n(n+1)}\;,\;\;\;\;
\delta\gamma_{gq}^{(0)n} = - 4C_F\, \frac{n+2}{n(n+1)}\\
\delta\gamma_{gg}^{(0)n} & = & 4 C_A \left[2 S_1(n) -\frac{4}{n(n+1)}
-\frac{11}{6}\right] + \frac{8}{3} T_f
\end{eqnarray}
where $C_F=4/3$, $C_A=3$ and $T_f=f/2$, with $f$ being the number of active
flavors ($f=3$ has been used when calculating $\delta \gamma_{ij}$).
Note that $\delta\gamma_{NS}^{(0)n}=\delta\gamma_{qq}^{(0)n}=\gamma_{qq}^{(0)n}$
where the latter quantity refers to the spin-averaged (unpolarized)
anomalous dimension. Furthermore, for the first $n=1$ moment we have
$\delta \gamma_{qq}^{(0)1}=\delta \gamma_{qg}^{(0)1}=0$ as a consequence of
helicity conservation at the quark-gluon vertex.

The spin-dependent NLO ($\overline{\rm{MS}}$) two-loop flavor non-singlet
anomalous dimensions $\delta\gamma_{NS}^{(1)n}(\eta)$, required in (2.10)
for the evolution of $\delta q_{NS\,\eta=\pm}^n(Q^2)$,
are the same as found for the spin-averaged case, $\delta\gamma_{NS}^{(1)n}
(\eta)=\gamma_{NS}^{(1)n}(\eta)$ with $\gamma_{NS}^{(1)n}(\eta=\pm 1)$
being given by eq.(B.18) of \cite{ref36}. Note that
$\delta\gamma_{NS}^{(1)n}(\eta=+1)$ governs the evolution of the NS
combinations $\delta q - \delta \bar{q}$, while
$\delta\gamma_{NS}^{(1)n}(\eta=-1)$ refers to the combinations
$\delta q +\delta\bar{q}$ appearing in the NS expressions (2.11).
The NLO flavor singlet anomalous dimensions $\delta\gamma_{ij}^{(1)n}$ in
the $\overline{\rm{MS}}$ scheme are as follows \cite{ref11}:
\begin{equation}
\delta\gamma_{qq}^{(1)n} =\gamma_{NS}^{(1)n}(\eta=-1) + \delta
\gamma_{PS,qq}^{(1)n}
\end{equation}
with $\gamma_{NS}^{(1)n}(\eta=-1)$ being again given by eq.(B.18) of
\cite{ref36} and\footnote{
Note that $\delta P_{ij}^{(0,1)}(x)\rightarrow P_{ij}^{(0,1)}(x)$ as
$x\rightarrow 1$ (or equivalently $\delta \gamma_{ij}^{(0,1)n}\rightarrow
\gamma_{ij}^{(0,1)n}$ as $n\rightarrow \infty$) except for
$\delta P_{gq}^{(1)}(x)>P_{gq}^{(1)}(x)$ as $x\rightarrow 1$. This
latter property has, however, no quantitative consequences for the
positivity, eq.(1.3), of our resulting NLO parton distributions.}
\begin{equation}
\delta\gamma_{PS,qq}^{(1)n} = 16 C_F T_f\;
\frac{n^4+2 n^3 + 2 n^2 + 5 n + 2}{n^3 (n + 1)^3} \\
\end{equation}
\vspace*{-1.0cm}
\begin{eqnarray}
\nonumber
\delta\gamma_{qg}^{(1)n} &=& 8 C_F T_f \left[ 2\, \frac{n-1}{n (n+1)}
\left( S_2(n)- S_1^2(n) \right) + 4 \,
\frac{n-1}{n^2 (n+1)} S_1(n)  \right. \\
\nonumber
&& \left. - \frac{5 n^5+5 n^4 -10 n^3 -n^2 +3n -2}{n^3 (n+1)^3} \right] \\
\nonumber
&+& 16 C_A T_f \left[ \frac{n-1}{n (n+1)}
\left(-S_2(n)+S_2^{'}\left(\frac{n}{2}\right)+S_1^2(n)\right) - \frac{4}
{n(n+1)^2} S_1(n)  \right. \\
&& \left. - \frac{n^5+n^4-4n^3+3n^2-7n -2}{n^3(n+1)^3}\right]\\
\nonumber
\delta\gamma_{gq}^{(1)n} &=&
32 C_F T_f \left[- \frac{n+2}{3n(n+1)} S_1(n) +\frac{5 n^2+12 n+ 4}
{9 n (n+1)^2} \right] \\
\nonumber
&+& 4 C_F^2 \left[2\, \frac{n+2}{n (n+1)} \left(S_2(n)+S_1^2(n)\right)-
2\, \frac{3 n^2+7 n +2}{n (n+1)^2} S_1(n) \right. \\
\nonumber
&&\left. +\frac{9n^5+ 30 n^4+24 n^3-7 n^2 - 16 n -4}{n^3 (n+1)^3}
\right] \\
\nonumber
&+& 8 C_A C_F \left[ \frac{n+2}{n (n+1)} \left(-S_2(n)+S_2^{'}\left(
\frac{n}{2}\right)-S_1^2(n)\right)+
\frac{11 n^2+ 22 n+ 12}{3n^2 (n+1)} S_1(n)  \right. \\
&& \left. - \frac{76 n^5 + 271 n^4 + 254 n^3 + 41 n^2 + 72 n +36}
{9 n^3 (n+1)^3} \right]\\
\nonumber
\delta\gamma_{gg}^{(1)n} &=&
8 C_F T_f\; \frac{n^6+3 n^5+ 5 n^4+ n^3-8 n^2+2 n+ 4}{n^3 (n+1)^3}\\
\nonumber
&+& 32 C_A T_f \left[-\frac{5}{9} S_1(n) +
\frac{3 n^4+6 n^3 + 16 n^2 + 13 n - 3}{9 n^2 (n+1)^2}\right] \\
\nonumber
&+& 4 C_A^2 \left[-S_3^{'}\left(\frac{n}{2}\right)-4 S_1(n) S_2^{'}
\left(\frac{n}{2}\right) + 8 \tilde{S}(n) +\frac{8}{n (n+1)}
S_2^{'}\left(\frac{n}{2}\right) \right. \\
\nonumber
&& \left. + 2 \,\frac{67 n^4 + 134 n^3 + 67 n^2 + 144 n + 72}
{9 n^2 (n+1)^2} S_1(n) \right. \\
&& \left. - \frac{48 n^6 + 144 n^5 + 469 n^4 + 698 n^3 + 7 n^2 + 258 n
+144}{9 n^3 (n+1)^3}\right]
\end{eqnarray}
where
\begin{eqnarray}
\nonumber
S_k(n) & \equiv & \sum_{j=1}^n \frac{1}{j^k}\\
\nonumber
S_k'\left(\frac{n}{2}\right) & \equiv & 2^{k-1} \sum_{j=1}^n
\frac{1+(-)^j}{j^k}\\
\nonumber
&=& \frac{1}{2} (1+\eta ) S_k\left(\frac{n}{2}\right)+
\frac{1}{2} (1-\eta ) S_k\left(\frac{n-1}{2}\right)\\
\nonumber
\tilde{S}(n) & \equiv & \sum_{j=1}^n \frac{(-)^j}{j^2} S_1(j)\\
& = & -\frac{5}{8} \zeta (3) +\eta \left[ \frac{S_1(n)}{n^2} +
\frac{\zeta (2)}{2} G(n) +\int_0^1 dx\; x^{n-1} \frac{{\rm{Li}}_2(x)}
{1+x}\right]
\end{eqnarray}
with $G(n)\equiv \psi\left(\frac{n+1}{2}\right) -
\psi\left(\frac{n}{2}\right)$
and $\eta\equiv (-)^n\rightarrow \pm 1$ for $\delta\gamma_{NS}^{(1)n}(
\eta=\pm 1)$ and $\eta \rightarrow -1 $ for the flavor singlet anomalous
dimensions (evolutions). The analytic continuations in $n$, required for the
Mellin inversion of these sums to Bj\o rken-$x$ space, are well known
\cite{ref16}.

It should be noted that the original results for $\delta\gamma_{ij}^{(1)n}$
have been presented \cite{ref11} in terms of multiple sums,
denoted by $\tilde{S}_k(n)$, $S_{k,l}(n)$ and $\tilde{S}_{k,l}(n)$, which
cannot be directly analytically continued in $n$. The following relations
have been used in order to arrive at (A.4)-(A.6):
\begin{eqnarray}
\nonumber
\tilde{S}_k(n) & \equiv & \sum_{j=1}^n \frac{(-)^j}{j^k}\\
\nonumber
& = & \frac{1}{2^{k-1}} S_k' \left( \frac{n}{2}\right) - S_k(n)\\
\nonumber
S_{1,2}(n)+S_{2,1}(n) & \equiv & \sum_{i=1}^n \left[ \frac{1}{i} S_2(i)+
\frac{1}{i^2} S_1(i)\right]\\
\nonumber
& = & S_1(n) S_2(n) + S_3(n)\\
\nonumber
\tilde{S}_{1,2}(n) & \equiv & \sum_{i=1}^n \frac{1}{i} \tilde{S}_2(i)\\
& = & S_1(n) \tilde{S}_2(n) +\tilde{S}_3(n) - \tilde{S}(n)
\end{eqnarray}
where for the latter sum we have used the identity
\begin{displaymath}
\sum_{i=1}^n \sum_{j=1}^i a_{ij} = \sum^n_{i=1} \sum_{j=1}^n a_{ij} -
\sum^n_{j=1} \sum^{j-1}_{i=1} a_{ij}
\end{displaymath}
in order to relate $\tilde{S}_{1,2}$ to the expressions in (A.7).

Finally, the first $n=1$ moments of the $q\rightarrow q(\bar{q})$ and
$g\rightarrow q(\bar{q})$ transitions reduce to \cite{ref11,ref37}
\begin{equation}
\delta\gamma_{NS}^{(1)1}(\eta = -1)=0\;,\;\;\;
\delta\gamma_{qq}^{(1)1}= 24 C_F T_f\;,\;\;\;
\delta\gamma_{qg}^{(1)1}=0\;\;\;.
\end{equation}
\newpage
\section*{Figure Captions}
\begin{description}
\item[Fig.1] Comparison of our NLO and LO results for
$A_1^{N} (x,Q^2) $ as obtained
from the fitted inputs at $Q^2=\mu^2_{NLO,LO}$ for the 'standard'
(eqs.(3.7) and (3.7')) scenario with present data [4-9].
The $Q^2$ values adopted here correspond to the different values quoted in
[4-9] for each data point starting at
$Q^2\,$\raisebox{-1mm}{${\stackrel{\textstyle >}{\sim}}$}$\,
1\,{\rm{GeV}}^2$ at the
lowest available $x$-bin. The results in the 'valence' scenario
are indistinguishable from the ones shown.
\item[Fig.2] The $Q^2$-dependence of $A_1^{p,d}(x,Q^2)$, as predicted by
the NLO and LO QCD evolutions at various fixed values of $x$,
compared with recent SLAC-E143 \cite{ref6} (solid circles) and
SMC data \cite{ref5,ref7} (open circles).
\item[Fig.3] The $Q^2$-dependence of $A_1^{p,n}(x,Q^2)$ as predicted by
the NLO and LO QCD evolutions at various fixed values of $x$.
\item[Fig.4a] Comparison of our 'standard' and 'valence' NLO and LO results
with the data [4-9] for
$g_1^{p,d}(x,Q^2)$. The SMC data correspond to different
$Q^2\,$\raisebox{-1mm}{${\stackrel{\textstyle >}{\sim}}$}$\,
1\,{\rm{GeV}}^2$ for $x\ge 0.005$, as do the theoretical results.
\item[Fig.4b] Same as in Fig.4a but for $g_1^n(x,Q^2)$. The E142 and
E143 data \cite{ref8,ref9} correspond to an average
$\langle Q^2\rangle =2$ and $3\,{\rm{GeV}}^2$, respectively,
and the theoretical
predictions correspond to a fixed $Q^2=3\,{\rm{GeV}}^2$.
\item[Fig.5a] Comparison of our fitted 'standard' and 'valence' input
NLO ($\overline{\rm{MS}}$) densities in eqs.(3.7) and (3.8) with the
unpolarized dynamical input densities of ref.\cite{ref14}.
\item[Fig.5b] The same as in Fig.5a but for the LO input densities according
to eqs.(3.7') and (3.8').
\item[Fig.6] The polarized LO and NLO ($\overline{\rm{MS}}$) densities at
$Q^2=4\,{\rm{GeV}}^2$, as obtained from the input densities at
$Q^2=\mu^2_{NLO,LO}$ in Figs.5a,b. In the 'standard' scenario, $\delta s$
coincides with the curves shown for $\delta\bar{q}$ in NLO and LO due to
the $SU(3)_f$ symmetric input which is only marginally broken in NLO
for $Q^2>\mu^2_{NLO}$.
\item[Fig.7] The experimentally allowed range of NLO polarized gluon
densities at $Q^2=4\,{\rm{GeV}}^2$ for the 'valence' scenario with
differently chosen $\delta g(x,\mu^2_{NLO})$ inputs. The 'fitted
$\delta g$' curve is identical to the one in Fig.6 and corresponds
to $\delta g(x,\mu^2_{NLO})$ in eq.(3.8). Very similar results are
obtained if $\delta g(x,\mu^2_{NLO})$ is varied accordingly within
the 'standard' scenario as well as in a LO analysis (see, e.g.,
ref.\cite{ref1}).
\end{description}
\newpage
\pagestyle{empty}

\vspace*{-2.1cm}
\hspace*{-0.7cm}
\epsfig{file=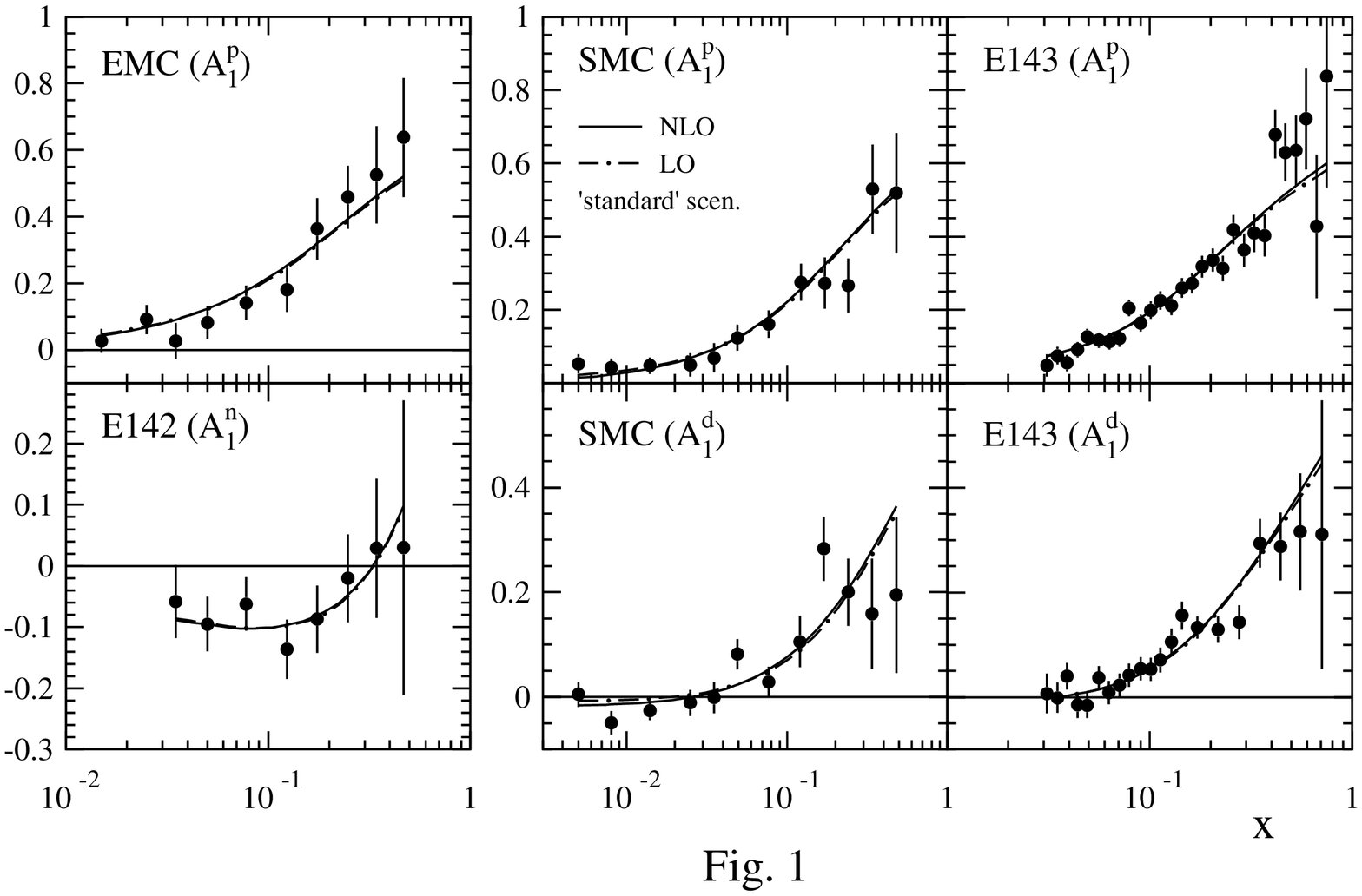,angle=90}

\newpage
\vspace*{-2cm}
\hspace*{-1.7cm}
\epsfig{file=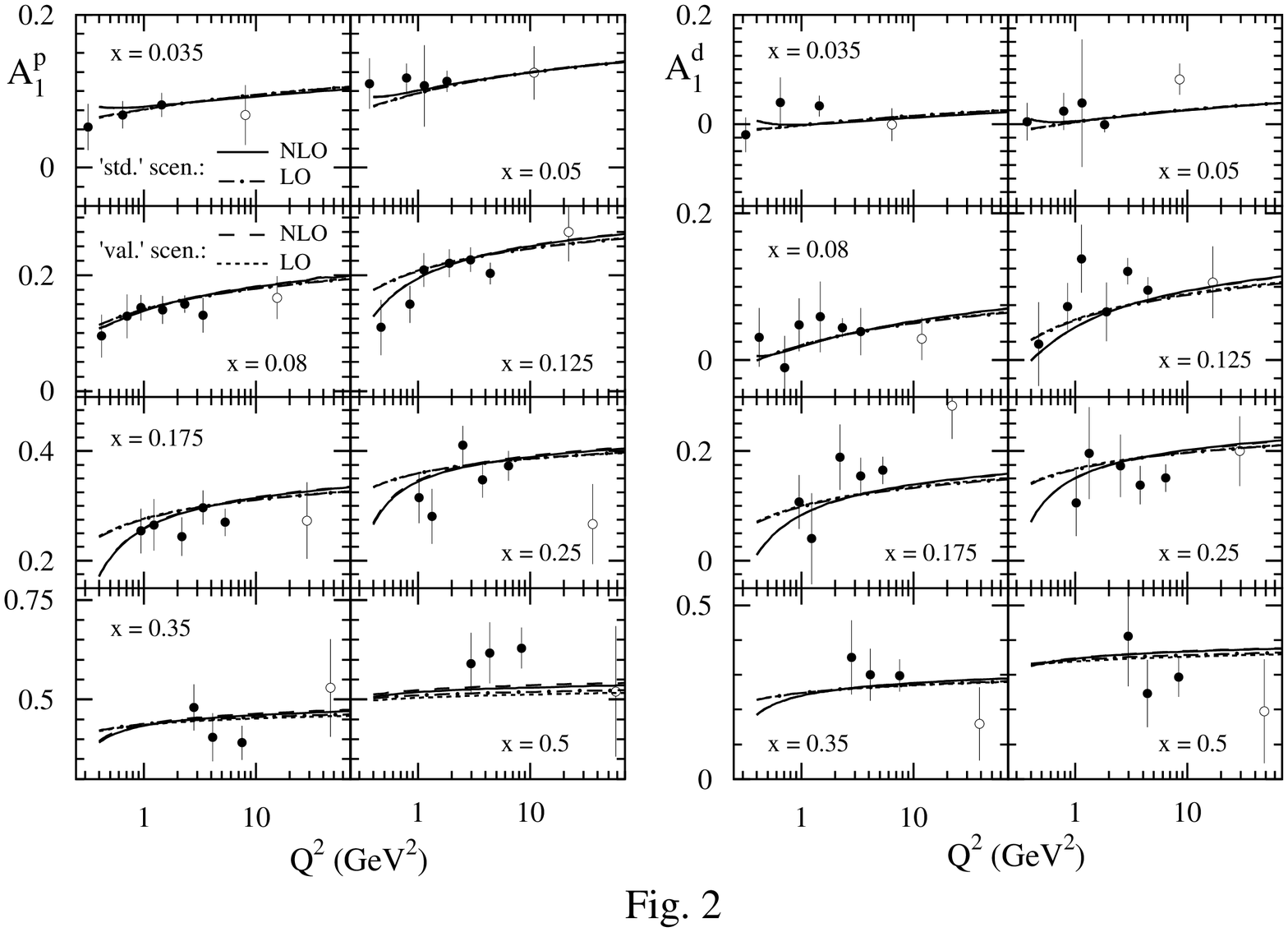,angle=90}

\newpage
\vspace*{-2cm}
\hspace*{-1.7cm}
\epsfig{file=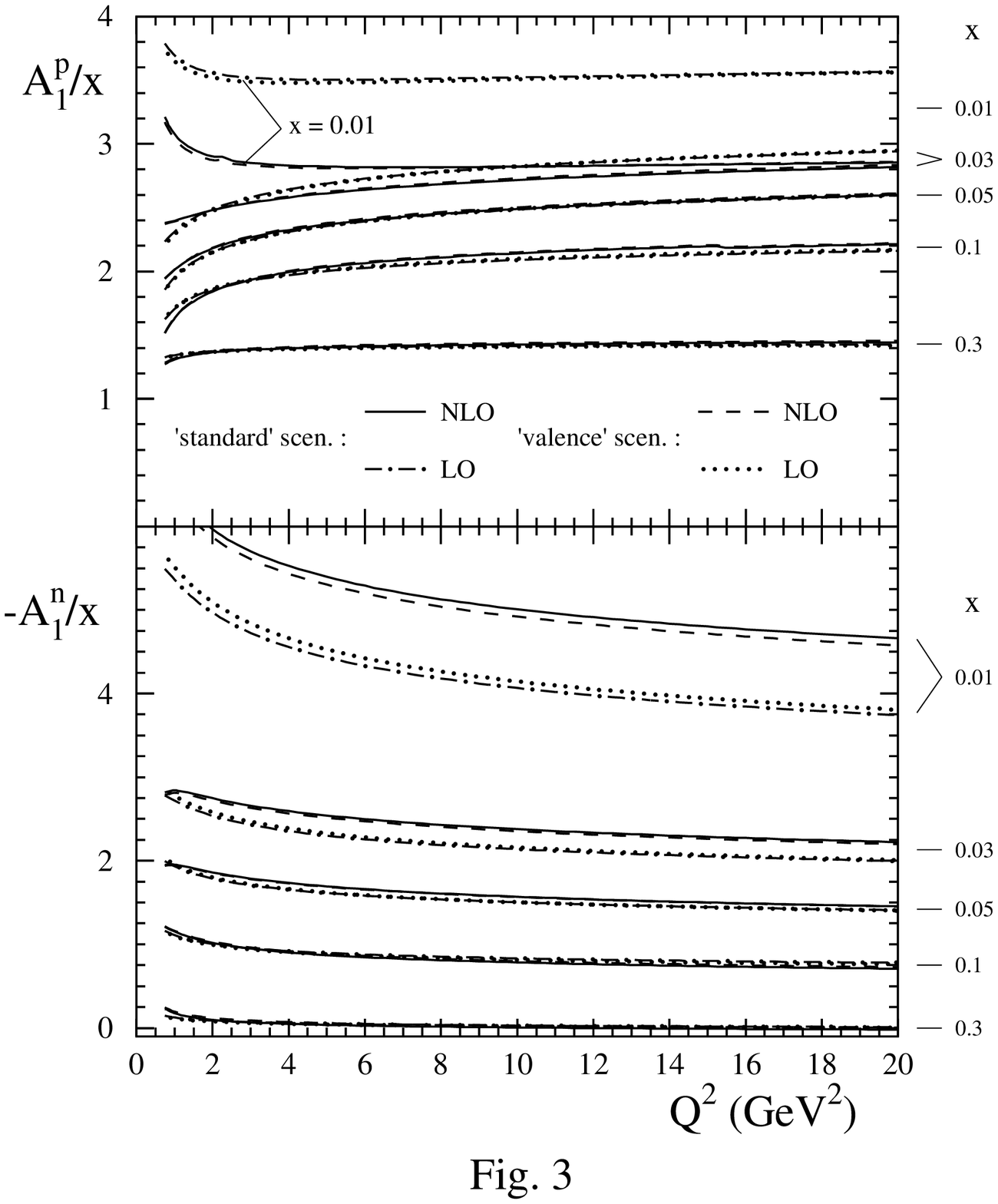}

\newpage
\vspace*{-1cm}
\hspace*{-0.7cm}
\epsfig{file=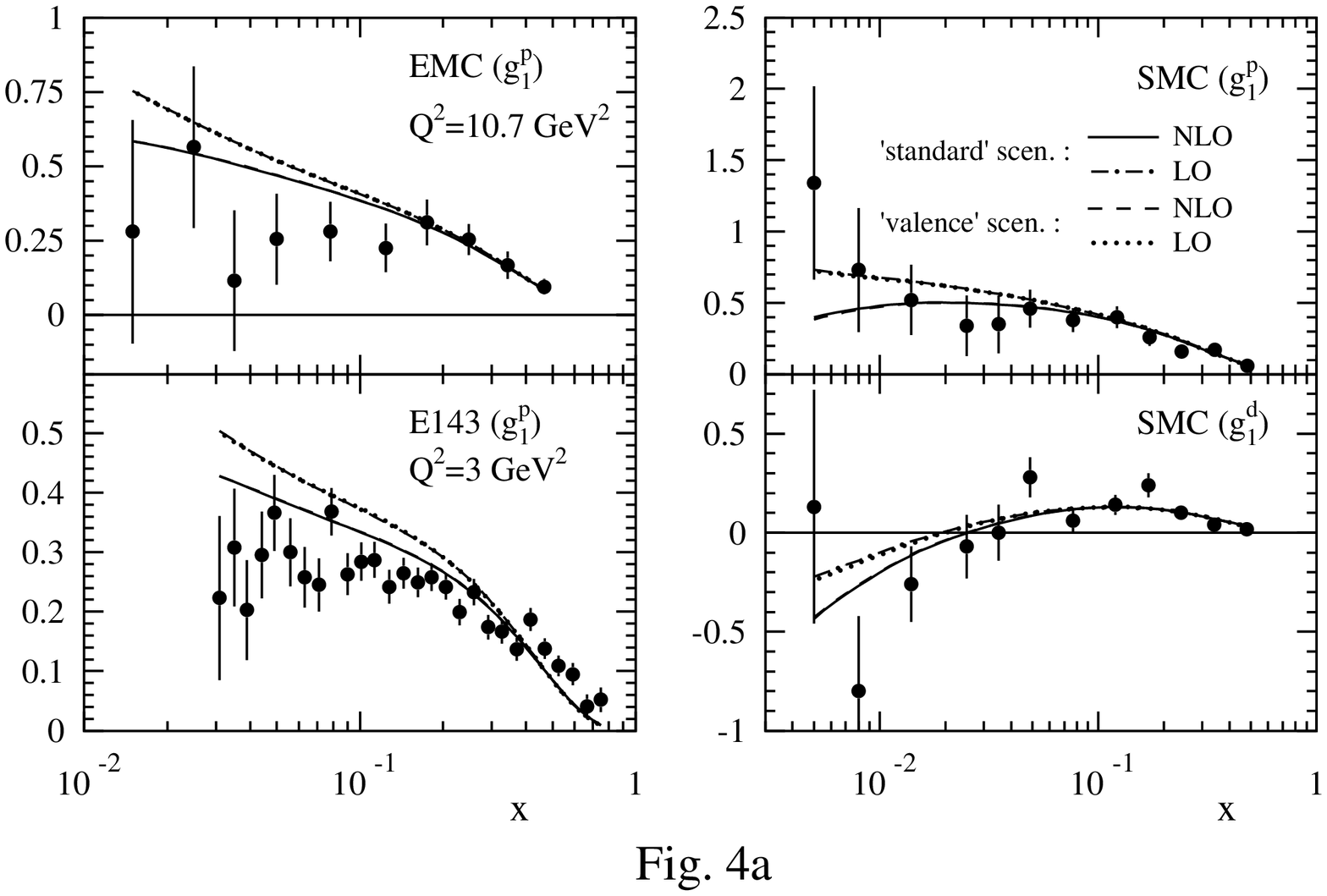,angle=90}

\newpage
\vspace*{0cm}
\hspace*{-1.3cm}
\epsfig{file=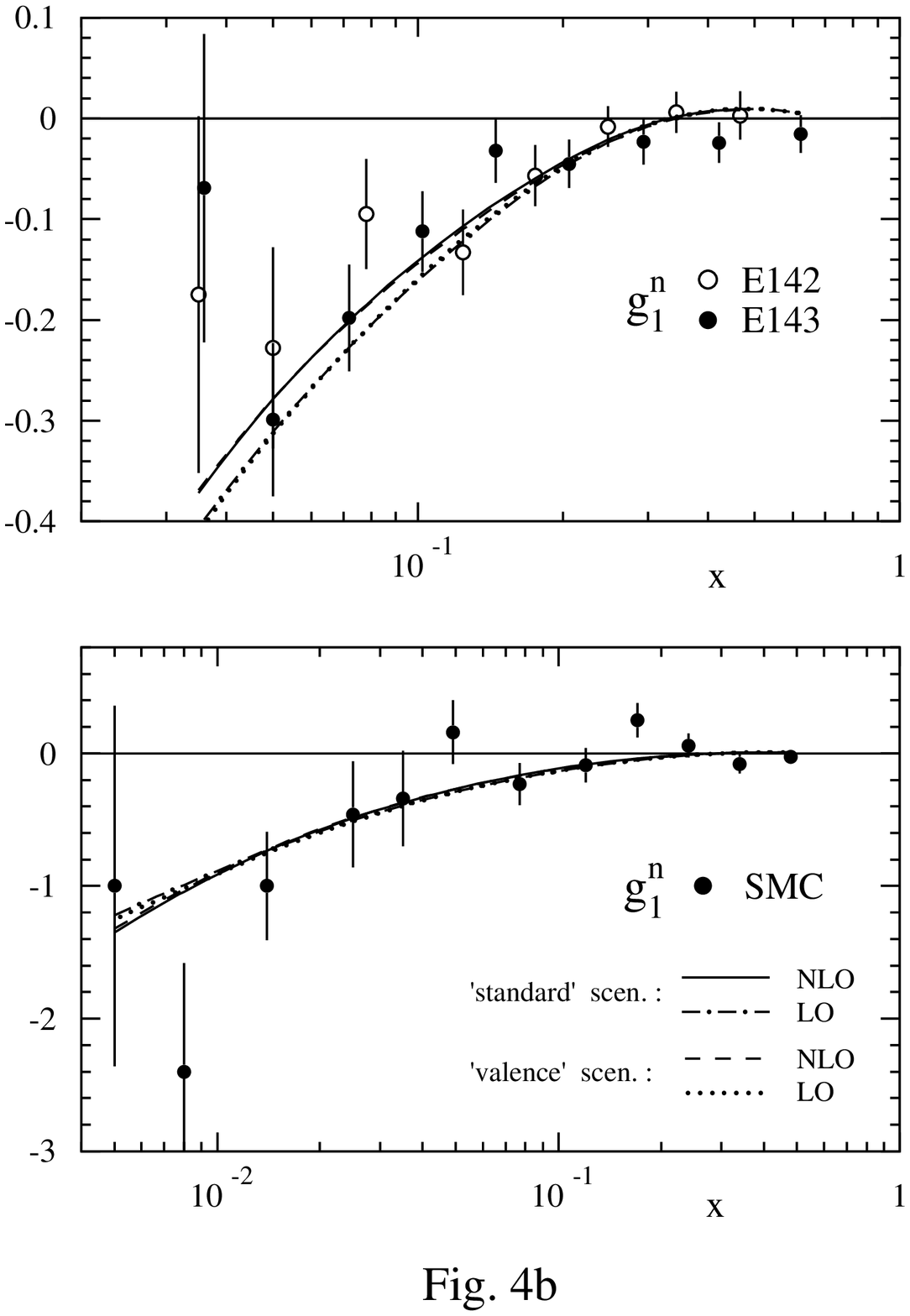}

\newpage
\vspace*{-2cm}
\hspace*{-0.7cm}
\epsfig{file=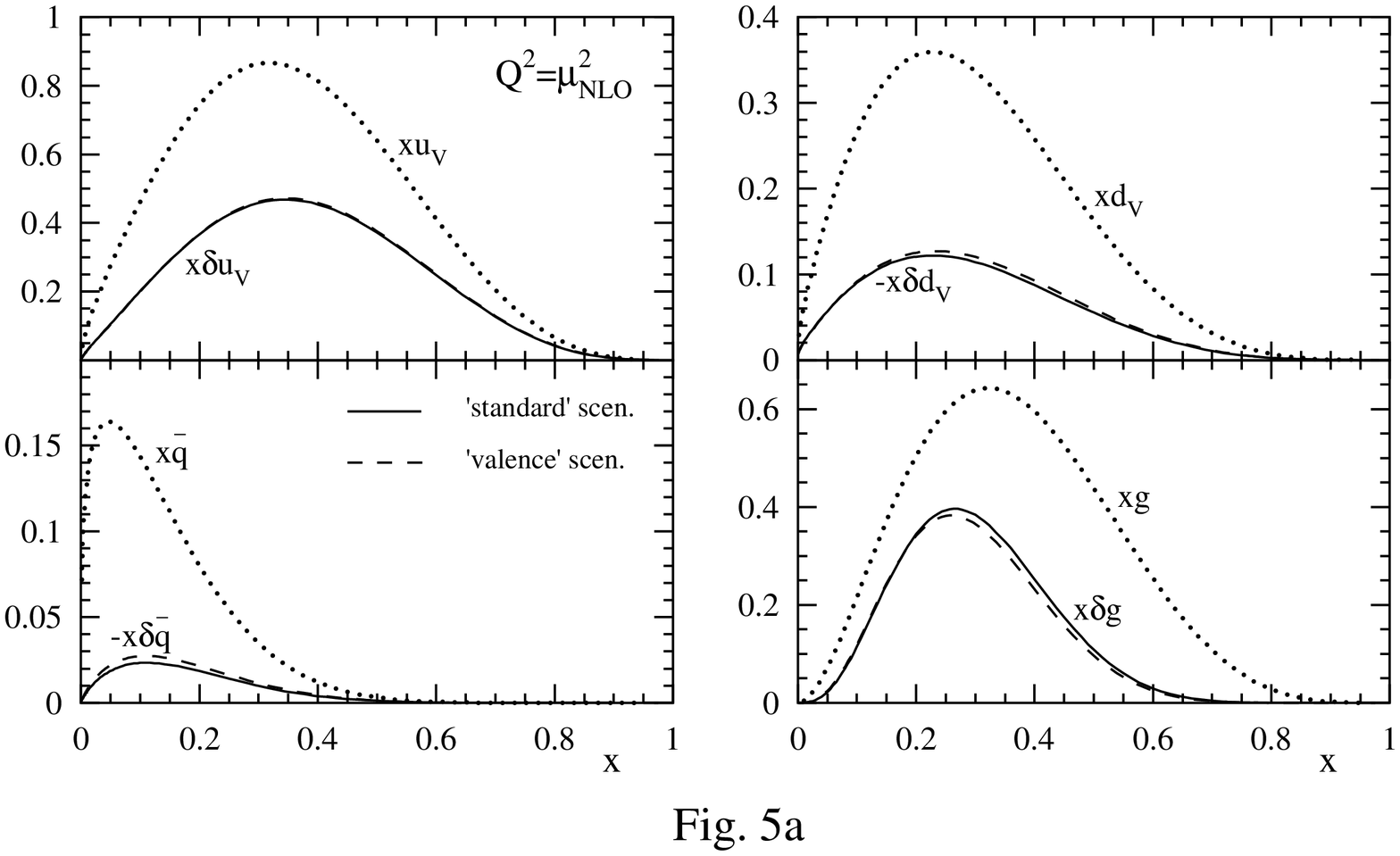,angle=90}

\newpage
\vspace*{-1cm}
\hspace*{-0.7cm}
\epsfig{file=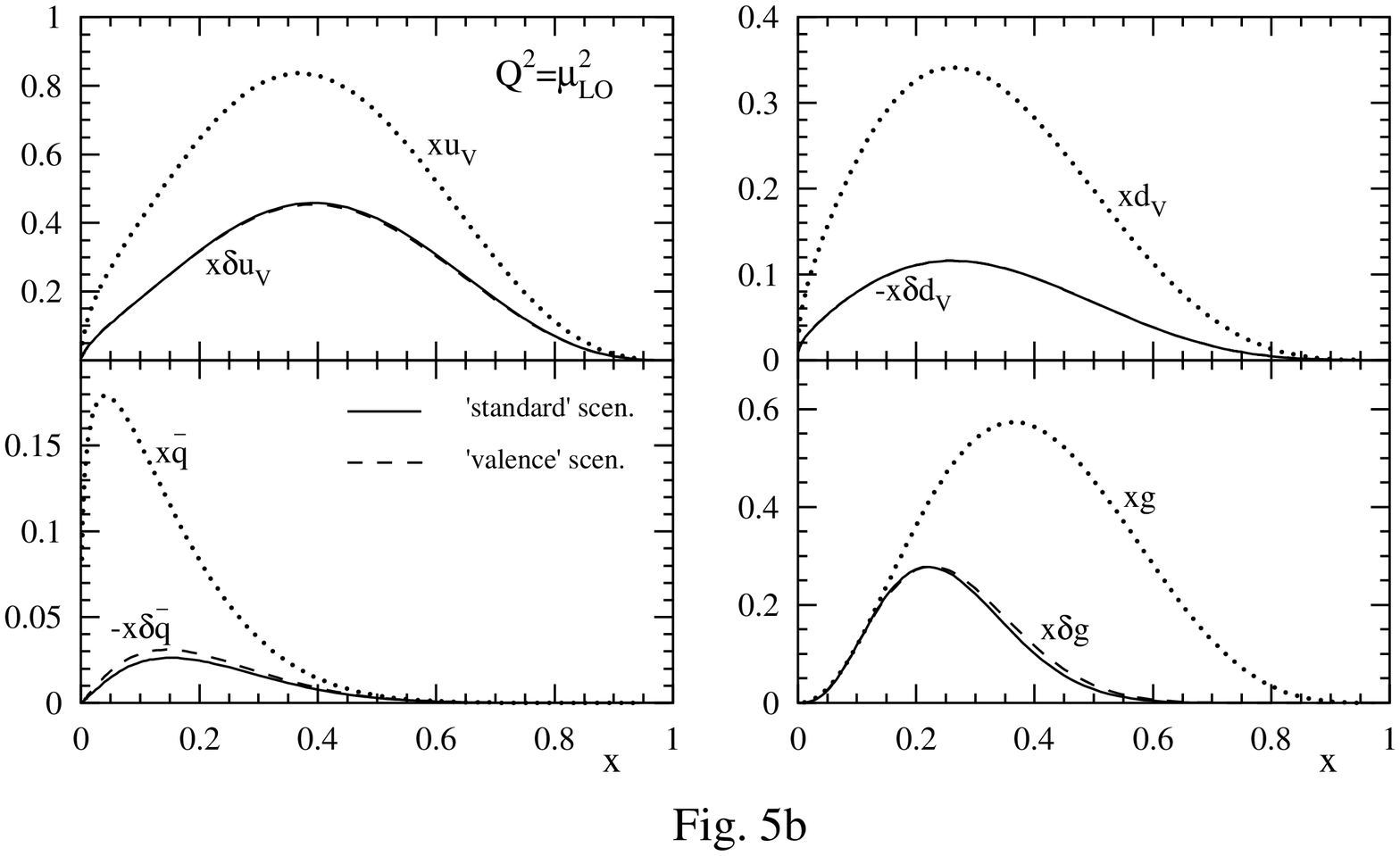,angle=90}

\newpage
\vspace*{-1cm}
\hspace*{-0.7cm}
\epsfig{file=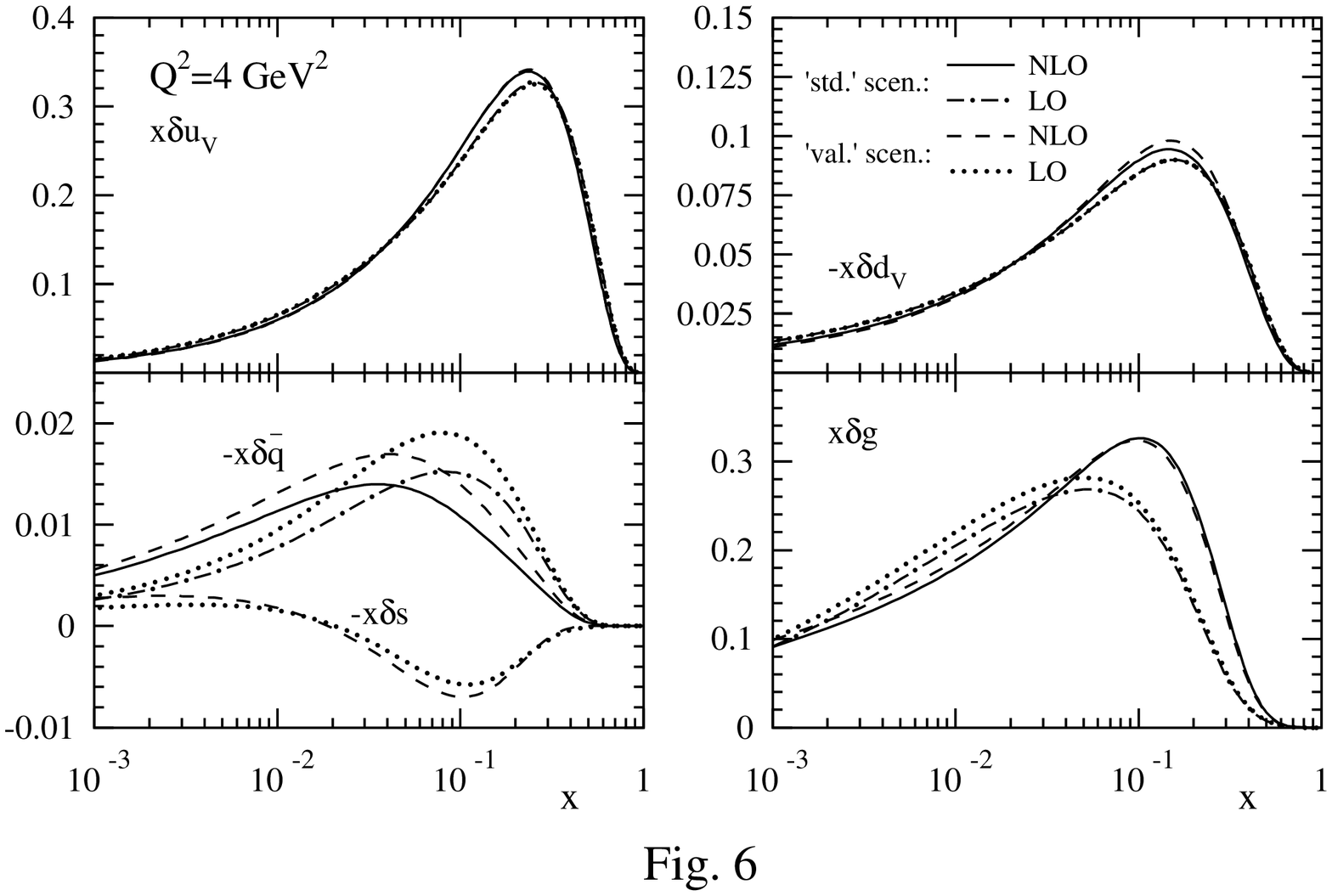,angle=90}

\newpage
\vspace*{0cm}
\hspace*{-1.4cm}
\epsfig{file=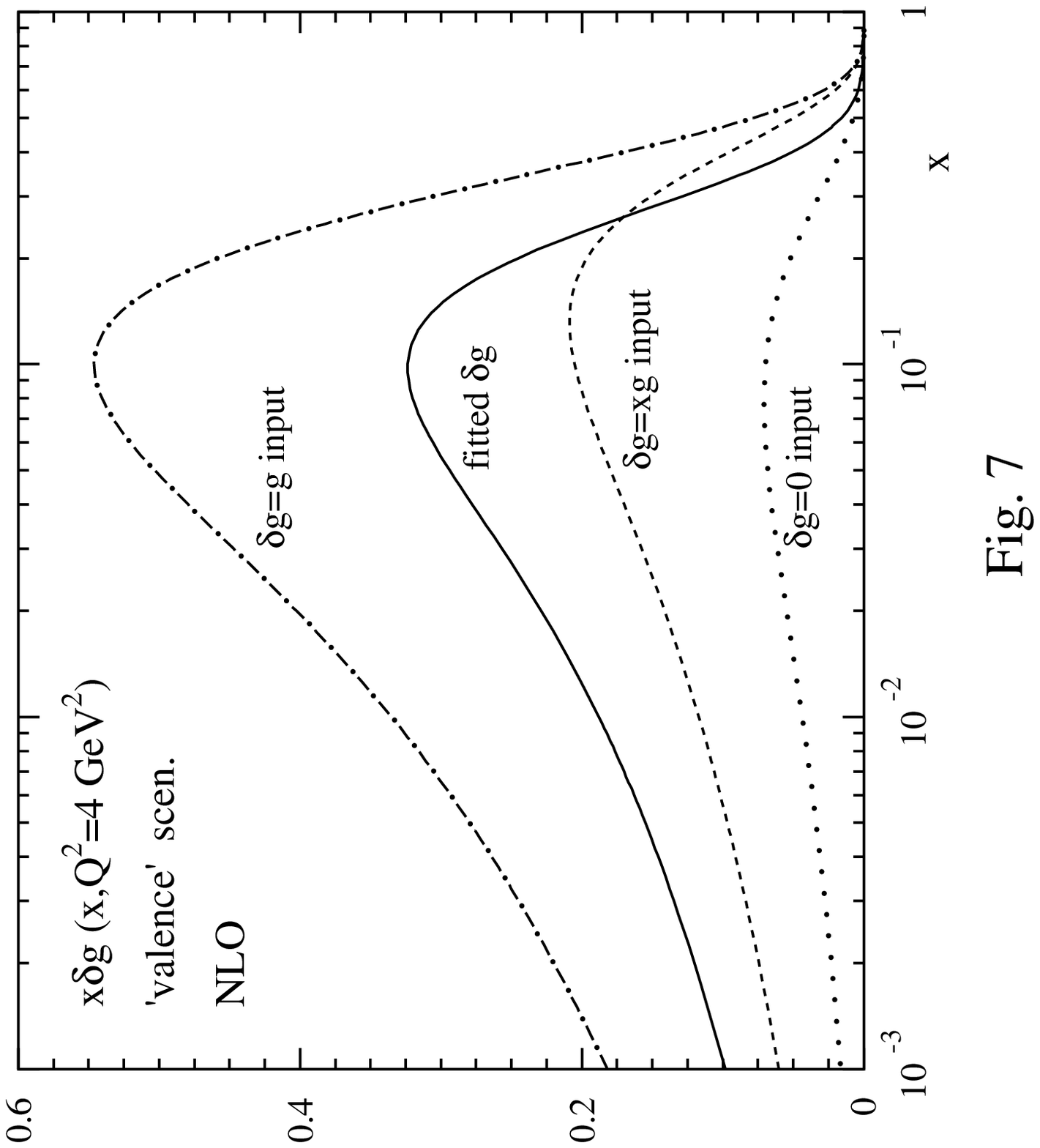,angle=90}

\end{document}